\documentclass[reprint,aps,prd,nofootinbib,superscriptaddress]{revtex4-1}
\usepackage{amsmath,amssymb,amsfonts}
\usepackage{graphicx}
\usepackage{graphics}
\usepackage{subfigure}
\usepackage{booktabs}
\usepackage{array}

\graphicspath{{figures/}}

\begin{document}

\title{Quasinormal modes of charged de Sitter black holes and strong cosmic censorship in conformal Killing gravity}

\author{Gao-Ming Deng}
\email{gmdeng2010@126.com}
\affiliation{School of Physics and Astronomy, China West Normal University, Sichuan 637002, China}

\author{Zong-Yao Zhang \thanks{Corresponding author}}
\email{zhangzongyao2026@126.com}
\affiliation{School of Physics and Astronomy, China West Normal University, Sichuan 637002, China}

\author{Yiwen Liu}
\affiliation{School of Physics and Astronomy, China West Normal University, Sichuan 637002, China}

\author{Si-Yuan Pan}
\affiliation{School of Physics and Astronomy, China West Normal University, Sichuan 637002, China}

\begin{abstract}
The strong cosmic censorship conjecture argues that the future evolution of the spacetime should be predictable from the initial data. However, within the framework of general relativity, some counterexamples to the conjecture have been found, especially for de Sitter spacetime. Spurred by challenges to the validity of the strong cosmic censorship conjecture, we extend such investigations to conformal Killing gravity, a promising candidate beyond Einstein's theory. Considering neutral massless scalar perturbations for charged de Sitter black hole, we not only shed light on the quasinormal mode spectrum, but also clarify whether the modifications from conformal Killing gravity and nonlinear electrodynamics can rescue the strong cosmic censorship conjecture, and further investigate how characteristic parameters affect its validity. As the first instance for investigating the conjecture in the framework of conformal Killing gravity, this paper reports some unexpected findings.
\end{abstract}

\maketitle

\section{Introduction} \label{Intro}

The strong cosmic censorship (SCC) conjecture has recaptured increasing attention recently. This conjecture, proposed by Penrose \cite{Penrose:1969pc}, asserts that the maximal Cauchy development of physically reasonable initial data in general relativity should be inextendible beyond the boundary known as Cauchy horizon. No observer can cross the Cauchy horizon into a spacetime region where the Einstein equations cease to determine the evolution uniquely. A widespread version of SCC is presented in Christodoulou's formulation \cite{Christodoulou:2008nj}, claiming that the spacetime metric cannot be extended beyond the Cauchy horizon as weak solutions to Einstein equations. To safeguard the determinism of classical general relativity, SCC should be respected in black hole spacetimes. It has been demonstrated that the SCC holds true for asymptotically flat black holes \cite{Dafermos:2012np,Luk:2015qja,Dafermos:2015bzz}. However, the fate of SCC becomes elusive for black holes in de Sitter spacetime. Cardoso \cite{Cardoso:2017soq} and Dias et al. \cite{Dias:2018etb} have successfully observed the SCC violation of (near-extremal) Reissner-Nordstr{\"o}m de Sitter black holes, while the SCC respect for the Kerr de Sitter case \cite{Dias:2018ynt}. Subsequently, a great many intriguing researches have explicitly discussed the SCC for various black hole geometries and perturbations \cite{Mo:2018nnu,Ge:2018vjq,Destounis:2018qnb,Deng:2018wrd,Dias:2019ery,Gan:2019jac,Liu:2019lon,Destounis:2019omd,Sang:2022frw,Shao:2023qlt,Jiang:2024cni,Davey:2024xvd}. It seems that the potential violation of SCC prevails for de Sitter black holes, especially for the nonrotating case. Interestingly, the validity of SCC exhibits a subtle sensitivity to the matter field, spacetime dimension, and the underlying gravitational dynamics. Can the confidence in SCC be restored when the gravitational and other modifications are taken into account? Motivated by this question, we extend the investigation of SCC to conformal Killing gravity, which emerges as a promising new candidate theory beyond Einstein's general relativity. As a highlight, it is the first instance of examining SCC in this fascinating extended theory.

Conformal Killing gravity (CKG), creatively introduced by Harada \cite{Harada:2023rqw}, presents a novel geometric picture in which gravitational dynamics are governed by the field equations augmented by a divergence-free conformal Killing tensor, effectively reducing the order of equations to the second. The solution space of CKG theory encompasses all solutions of general relativity. Remarkably, this theory satisfies several theoretical criteria that other existing theories, including well-known general relativity, fail to simultaneously fulfill. Namely, (i) the cosmological constant emerges as an integration constant rather than being fixed a priori; (ii) the energy conservation law arises directly from the field equations rather than being assumed; (iii) conformally flat metrics need not be vacuum solutions. A notable cosmological consequence is that CKG offers a compelling explanation for the accelerated expansion of the Universe without invoking dark energy \cite{Harada:2023rqw,Harada:2023afu}. Nowadays, CKG theory has been widely applied in astrophysics and cosmology \cite{Mantica:2024sdy,Feng:2024rnh,Capozziello:2025kws,Ladghami:2026lwg,Praveena:2026iei}.

Nonlinear electrodynamics (NLED) has been advanced as a hot topic in classical and quantum physics. More recently, Junior et al. \cite{Junior:2025zgd} constructed the solutions of charged black holes in CKG coupled with NLED and demonstrated that the shadow sizes are consistent with observational data. Inspired by the foregoing triumphs of CKG theory, the quasinormal modes (QNMs) and the strong cosmic censorship (SCC) of the charged de Sitter black holes \cite{Junior:2025zgd} become highly anticipated and deserve in-depth investigations. Considering neutral massless scalar field perturbations, we not only shed light on the QNM spectrum, but also clarify whether the CKG and NLED modifications can rescue SCC, and further analyze the significant effects of these modified parameters. Our discussions provide a new reference for validating the validity of SCC beyond Einstein's theory.

This paper is organized as follows. In the next section, we briefly review the black hole spacetime background. We concentrate on investigating the QNM spectrum and deriving the criterion that marks the SCC respect or violation in Sec.~\ref{SecQNMs}. Section \ref{SecSCC} is devoted to numerically examining the fate of the SCC and analyzing the subtle effects of modified parameters. Finally, Section \ref{SecCon} is reserved for conclusions. For convenience, we set $8\pi G =c=1$ throughout this paper.

\section{Charged de Sitter black holes in conformal Killing gravity} \label{SecCKGbh}

In this paper, we consider a novel charged de Sitter black hole spacetime in the theory of CKG \cite{Junior:2025zgd}. Coupling the CKG theory with NLED, Junior et al. generalized the Maxwell Lagrangian density as,
\begin{equation}
    \mathcal{L}(F) = a_0 F^{f_k}+F,
    \label{eqCKGLag}
\end{equation}
in which the electromagnetic scalar $F=\frac{1}{4}F^{\mu\nu}F_{\mu\nu}$. The gravitational field equation for this theory can be rewritten as an Einstein-like equation containing a characteristic divergence-free conformal-Killing tensor $K_{\mu\nu}$ \cite{Harada:2023afu,Mantica:2024sdy}, that is
\begin{equation}
G_{\mu\nu}+\Lambda g_{\mu\nu}=T_{\mu\nu} +K_{\mu\nu},
\label{eqfield}
\end{equation}
in which $G_{\mu\nu}$$=$$R_{\mu\nu}-\frac{1}{2} R g_{\mu\nu}$ is the Einstein tensor.

Considering the static spherically symmetric ansatz, the line element of this spacetime is given by \cite{Junior:2025zgd}
\begin{equation}
ds^2 = -f(r)dt^2 + \frac{1}{f(r)}dr^2 + r^2 d\Omega_2^2,
\label{eqlineelement}
\end{equation}
where
\begin{equation}
f(r) = 1 - \frac{2M}{r} + \frac{a_0 2^{1-f_k} q^{2f_k} r^{2-4f_k}}{4f_k - 3} + \frac{q^2}{r^2} - \frac{\Lambda r^2}{3}.
\label{eqmetric}
\end{equation}
The line element describes the charged de Sitter black hole spacetime in CKG theory coupled with NLED. It is characterized by the mass parameter $M$, the charge parameter $q$, the positive cosmological constant and parameters $a_0$ and $f_k$. Of especial interest, this solution would reduce to the Schwarzschild when taking the limit $f_k \rightarrow 1$, $a_0 \rightarrow 0$, $\Lambda \rightarrow 0$, $q \rightarrow 0$.

The causal structure of the black hole spacetime above admits three distinct horizons, the Cauchy horizon at $r=r_-$, the event horizon at $r=r_h$ and the cosmological horizon at $r=r_c$, satisfying $r_- < r_h < r_c$.  The surface gravity can be derived as
\begin{equation}
\kappa_i = \frac{1}{2}\left|f'(r_i)\right|
\label{eqsurg}
\end{equation}
where the subscript $``i"$ corresponds to $\{-, h, c\}$ for each horizon.

\section{Scalar perturbations and quasinormal modes} \label{SecQNMs}

QNMs serve as a practical tool for testing stability. In this section, we concentrate on investigating the QNMs of the de Sitter black holes aforementioned under neutral massless scalar perturbations and further analyzing the internal stability.

Introducing neutral massless scalar field $\Psi$ propagating on the background geometry as the perturbations, the evolution behavior is governed by the Klein-Gordon equation
\begin{equation}
\square \Psi=0.
\label{eqKG}
\end{equation}
Considering the time translation invariance and symmetries of the spacetime, we decompose the field as
\begin{equation}
\Psi(t, r, \theta, \phi) = \sum_{l,m} e^{-i\omega t} Y_{lm}(\theta, \phi) \frac{\psi(r)}{r},
\label{eqdecom}
\end{equation}
in which $Y_{lm}(\theta, \phi)$ is the spherical harmonics. The equation of motion can be expressed as the Schr\"{o}dinger-like form
\begin{equation}
\frac{d^2\psi(r)}{dr_*^2} + \left[\omega^2 - V(r)\right] \psi(r) = 0,
\label{eqschr}
\end{equation}
where
\begin{equation}
V(r) = \frac{f(r)}{r^2}\left[ \ell(\ell+1) + r f'(r) \right],
\label{eqeffepote}
\end{equation}
denotes an effective potential for the scalar perturbation, and $r_*$ is the tortoise coordinate
\begin{equation}
dr_* =  \frac{dr}{f(r)}.
\label{eqtortoise}
\end{equation}

Physically, it is required that there only exists the pure ingoing wave near the event horizon and the pure outgoing wave near the cosmological horizon \cite{Konoplya:2011qq,Lin:2021fth}. Imposing the boundary conditions
\begin{equation}
\psi \left(r \to r_{h} \right) \sim e^{-i\omega r_*}, ~~~~ \psi \left(r \to r_{c}\right) \sim e^{i\omega r_*},
\label{eqbo}
\end{equation}
the spectrum of QNMs can be derived from the equation of motion \eqref{eqschr}. The eigenvalue appears as the discrete QNM frequency $\omega$.


To facilitate investigating the SCC, the extendibility of solutions should be further discussed. According to Christodoulou's definition of SCC, it requires that solutions to the equation of motion remains inextendible near the Cauchy horizon. Contrarily, if there exists a weak solution beyond the Cauchy horizon, the asymptotic solution can be expressed as \cite{Destounis:2019omd,Sang:2022frw}
\begin{equation}
\Psi \sim e^{i \frac{\omega}{\kappa_-} \ln|r - r_-|} \propto |r - r_-|^{\beta},
\label{eqasypsolution}
\end{equation}
in which we define
\begin{equation}
\beta := -\frac{\rm{Im}(\omega)}{\kappa_-},
\label{eqbeta}
\end{equation}
where $\rm {Im}(\omega)$ denotes the imaginary part of the lowest-lying/dominant, non-zero QNM frequency $\omega$. In fact, the extendibility of solutions near the Cauchy horizon for de Sitter spacetime is closely related to the competition between the exponential decay of perturbations and the blue-shift amplification. When perturbations decay exponentially fast enough to counterbalance the blue-shift effect, the solutions might be extendible beyond the Cauchy horizon, implying a potential violation of SCC. Therefore, $\beta$ can serve as a well-defined indicator to measure the validity of SCC.

For the scalar QNM solution extended beyond the Cauchy horizon, the local square integrability requires the radial contribution of scalar field
\begin{eqnarray}
\int_\mathcal{V} (\partial_r^2 {\Psi})^2dr &\sim& \int_\mathcal{V} |r-r_-|^{2(\beta-1)}dr \nonumber \\
                                           &=& \frac{|r-r_-|^{2\beta-1}}{2\beta-1}
\label{eqradial}
\end{eqnarray}
is finite near the Cauchy horizon, hence
\begin{equation}
\beta >\frac{1}{2}.
\label{eqcriteria}
\end{equation}
Notably, considering the neutral massless scalar perturbations, we obtain the crucial criterion to measure the validity of the SCC for the black hole spacetime in CKG theory coupled with NLED. If $\beta$$>$$1/2$ for all QNMs, the scalar field can extend across the Cauchy horizon as a weak solution, and Christodoulou's formulation of SCC might be violated. Conversely, the Cauchy horizon becomes inextendible when $\beta$$\leq$$1/2$, implying the SCC holds. It is noteworthy that the threshold for the criterion in the CKG theory takes the same value as that for Reissner-Nordstr{\"o}m de Sitter black holes in Einstein's theory \cite{Cardoso:2017soq,Liu:2019lon}.

\section{Strong cosmic censorship} \label{SecSCC}

In this section, we focus on numerically validating the SCC and systematically analyzing how characteristic parameters, such as the CKG and NLED modifications, influence its validity. With the criterion above in hand, we are interested in the values of $\beta$, especially for the dominant QNMs.

\subsection{Numerical methods} \label{Nummetho}

In this paper, we mainly employ the pseudospectral method \cite{Jansen:2017oag,Miguel:2020uln} to compute the QNMs. Meanwhile, the finite difference method \cite{Zhu:2014sya} and the WKB approximation \cite{Konoplya:2003ii,Deng:2014bta,Deng:2017bjz} are used to validate the reliability of our analyses.

The pseudospectral method discretizes the continuous differential problem by collocation, reducing it to a system of algebraic equations. To perform a pseudospectral grid discretization, we define the compact coordinate
\begin{equation}
u=\frac{r-r_h}{r_c-r_h},
\label{eqcompactcoor}
\end{equation}
which maps the original radial domain $r \in [r_h, r_c]$ within the finite interval $u \in [0, 1]$. $u = 0,1$ corresponds to the event and cosmological horizon respectively.

The singular power-law behavior near the two horizons can be described by
\begin{equation}
\psi(u) = u^{\frac{-i\omega}{f'(r_h)}}(1-u)^{\frac{i\omega}{f'(r_c)}}y(u),
\label{eqpowerlaw}
\end{equation}
where $y(u)$ remains regular within the interval $u\in[0,1]$. Substituting into the radial equation yields a second-order ordinary differential equation
\begin{equation}
\alpha_0(\omega,u)y(u)+\alpha_1(\omega,u)y'(u)+\alpha_2(\omega,u)y''(u)=0.
\label{eqradialode}
\end{equation}

Following the discussion in \cite{Jansen:2017oag} and introducing Chebyshev collocation points $u_j$,
the equation of motion can be discretized as a matrix form
\begin{equation}
\left(
M_0+\omega M_1+\omega^2M_2
\right)\mathbf y=0,
\label{eqmatrix}
\end{equation}
where the vector $\mathbf{y}$ is given by $\mathbf{y}_j$$=$$y(u_j)$. Eventually, the QNM frequencies can be computed by solving for the eigenvalues of the matrix equation.

\subsection{Numerical results} \label{Numresu}

In Tables \ref{tab:qnm_comparison_lambda_001} and \ref{tab:qnm_comparison_lambda_01}, we list some numerical results for the dominant QNMs obtained via three different methods. The good agreement of the data from the pseudospectral method with those from the other two cross-checking approaches supports the robustness of our analysis.
\begin{table*}[t]
\centering
\caption{The dominant QNMs with different charge ratio $q/q_{\rm{max}}$ calculated by three methods for fixed $l$. Here we set $\Lambda M^2=0.01$, $a_0=1.5$, $f_k=1.5$.}
\label{tab:qnm_comparison_lambda_001}
\scriptsize
\setlength{\tabcolsep}{5pt}
\begin{ruledtabular}
\begin{tabular}{lccccc}
$q/q_{\rm{max}}$         & $0.7$       & $0.8$       & $0.9$      & $0.99$     & $0.998$     \\ \hline
Pseudospectral method    & $0.093518$  & $0.093328$  & $0.091995$ & $0.087581$ & $0.086841$  \\
Finite difference method & $0.093477$  & $0.093130$  & $0.091853$ & $0.087741$ & $0.087004$  \\
WKB approximation        & $0.093518$  & $0.093283$  & $0.091968$ & $0.087577$ & $0.086781$  \\
\end{tabular}
\end{ruledtabular}
\end{table*}
\begin{table*}[t]
\centering
\caption{The dominant QNMs with different charge ratio $q/q_{\rm{max}}$ calculated by three methods for fixed $l$. Here we set $\Lambda M^2=0.1$, $a_0=1.5$, $f_k=1.5$.}
\label{tab:qnm_comparison_lambda_01}
\begingroup
\scriptsize
\setlength{\tabcolsep}{5pt}
\begin{ruledtabular}
\begin{tabular}{lccccc}
$q/q_{\rm{max}}$         & $0.7$       & $0.8$       & $0.9$      & $0.99$     & $0.998$  \\ \hline
Pseudospectral method    & $0.047234$  & $0.051408$  & $0.055127$ & $0.055464$ & $0.054999$      \\
Finite difference method & $0.047252$  & $0.051449$  & $0.055139$ & $0.055463$ & $0.055004$      \\
WKB approximation        & $0.047259$  & $0.051641$  & $0.055131$ & $0.055464$ & $0.054997$      \\
\end{tabular}
\end{ruledtabular}
\endgroup
\end{table*}

\begin{figure*}[htbp]
   \centering
   \subfigure{
   \begin{minipage}[h]{0.315\linewidth}
   \centering
   \includegraphics[width=\linewidth]{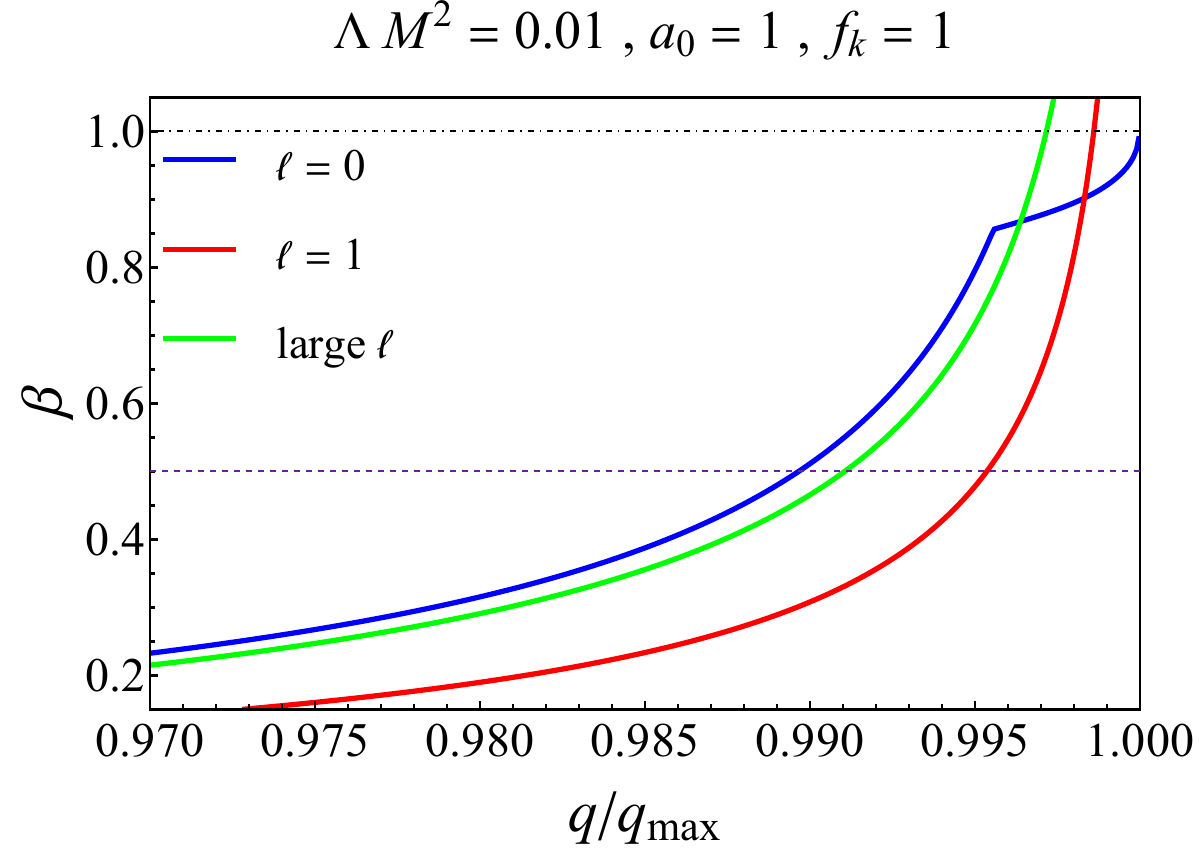}
   \end{minipage}
   }
    \subfigure{
   \begin{minipage}[h]{0.315\linewidth}
   \centering
   \includegraphics[width=\linewidth]{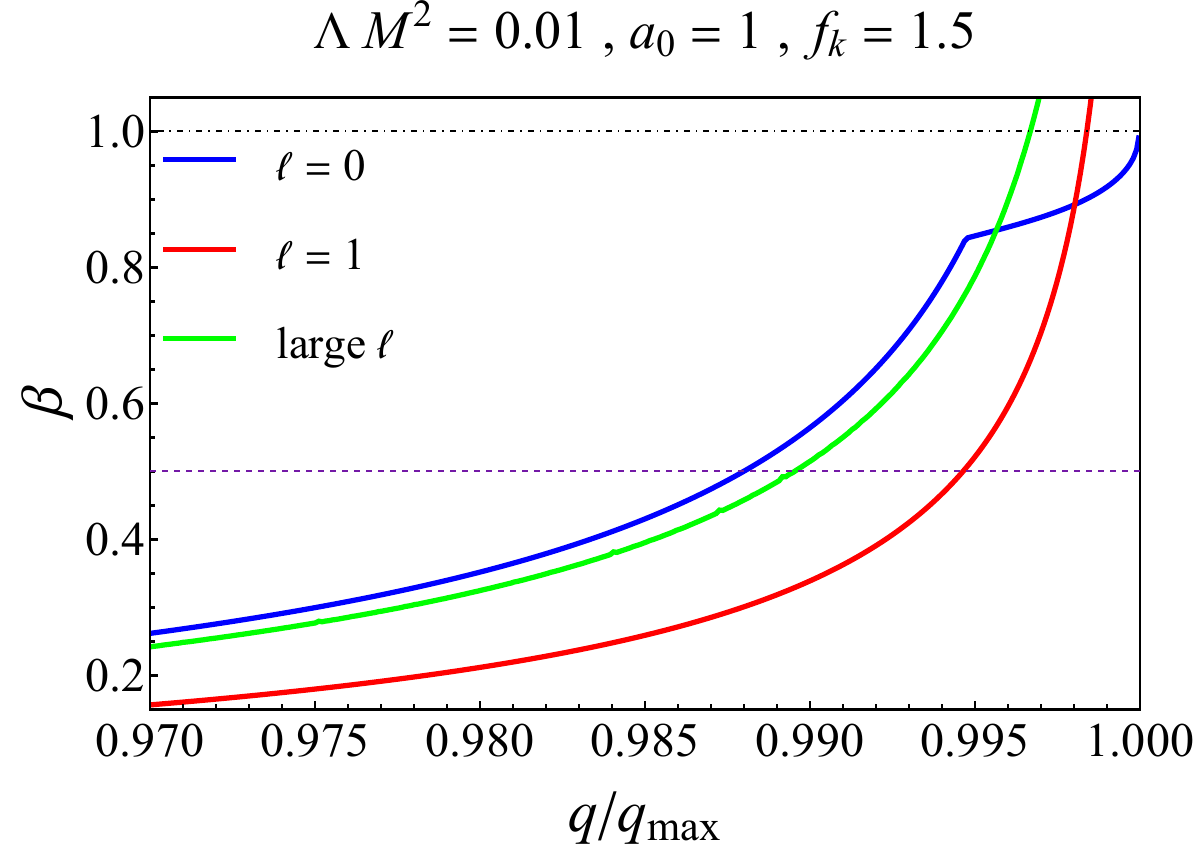}
   \end{minipage}
   }
   \subfigure{
   \begin{minipage}[h]{0.315\linewidth}
   \centering
   \includegraphics[width=\linewidth]{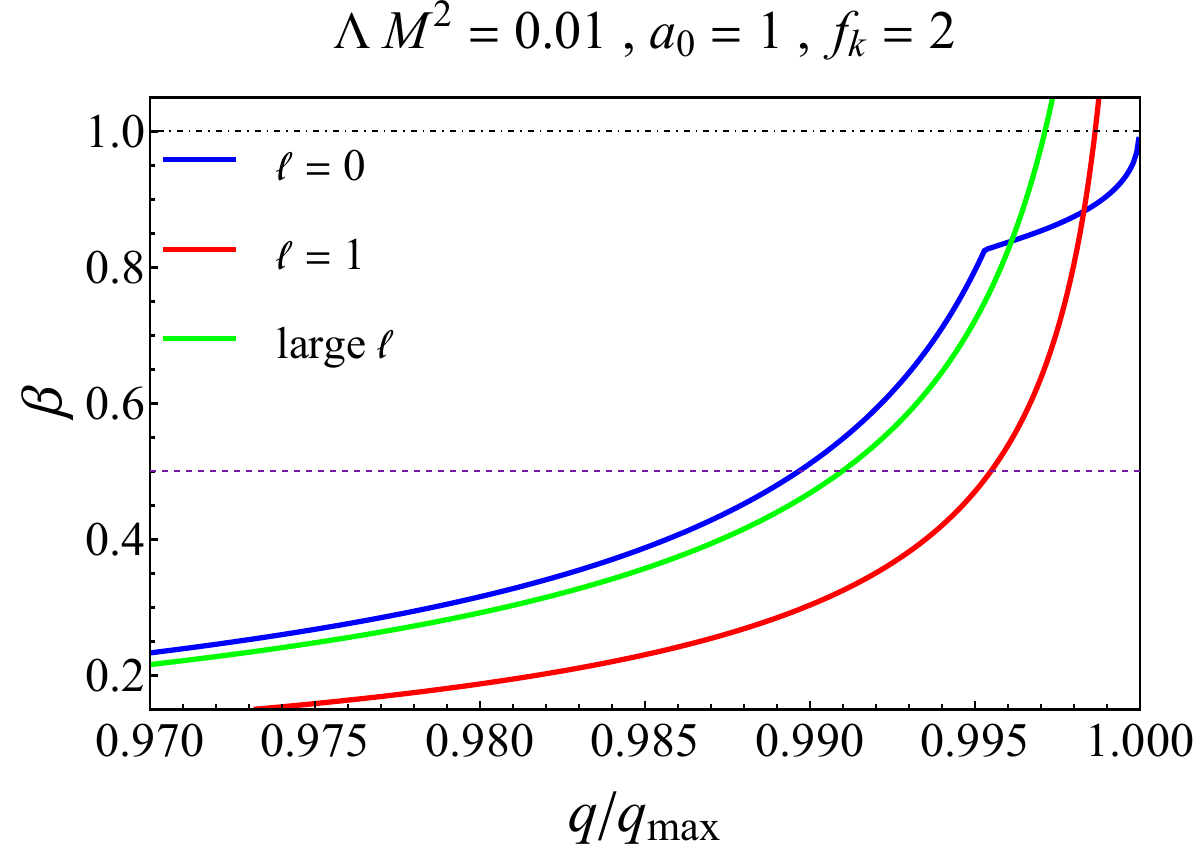}
   \end{minipage}
   }
    \subfigure{
   \begin{minipage}[h]{0.315\linewidth}
   \centering
   \includegraphics[width=\linewidth]{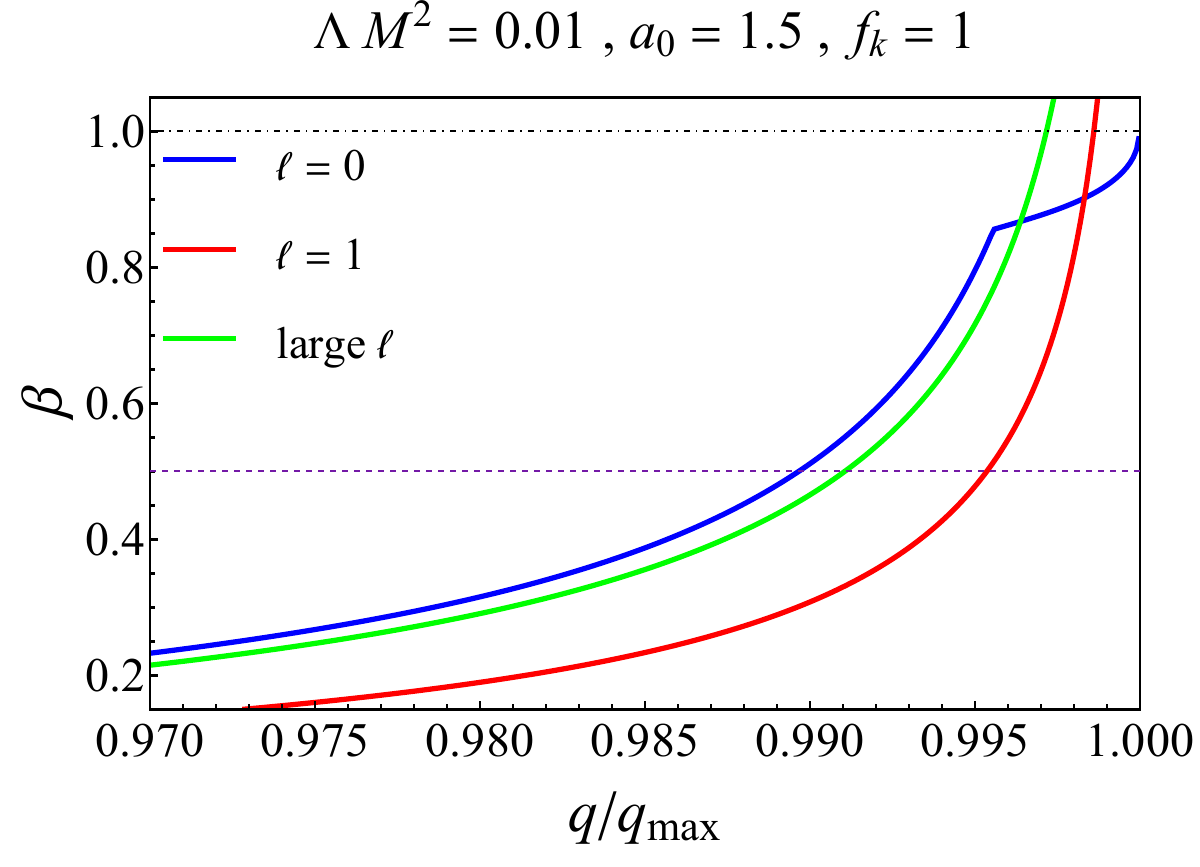}
   \end{minipage}
   }
    \subfigure{
   \begin{minipage}[h]{0.315\linewidth}
   \centering
   \includegraphics[width=\linewidth]{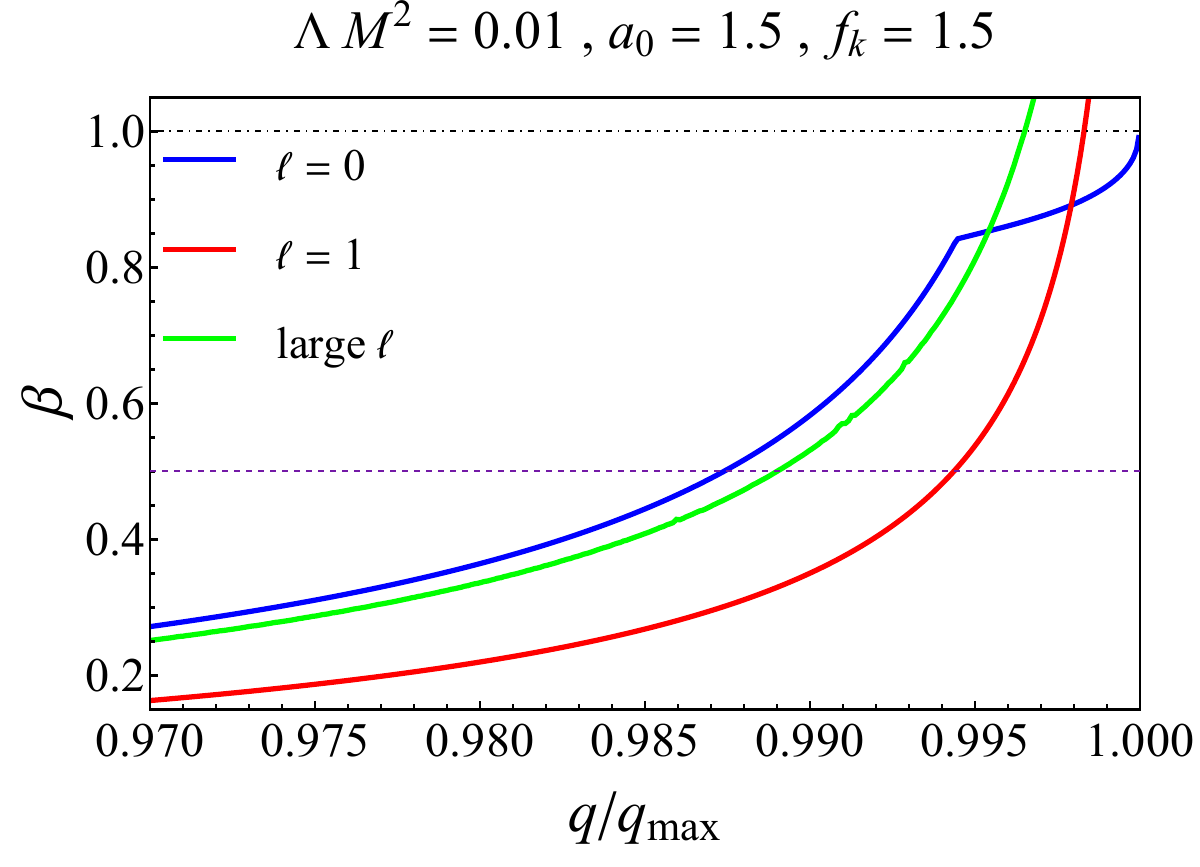}
   \end{minipage}
   }
   \subfigure{
   \begin{minipage}[h]{0.315\linewidth}
   \centering
   \includegraphics[width=\linewidth]{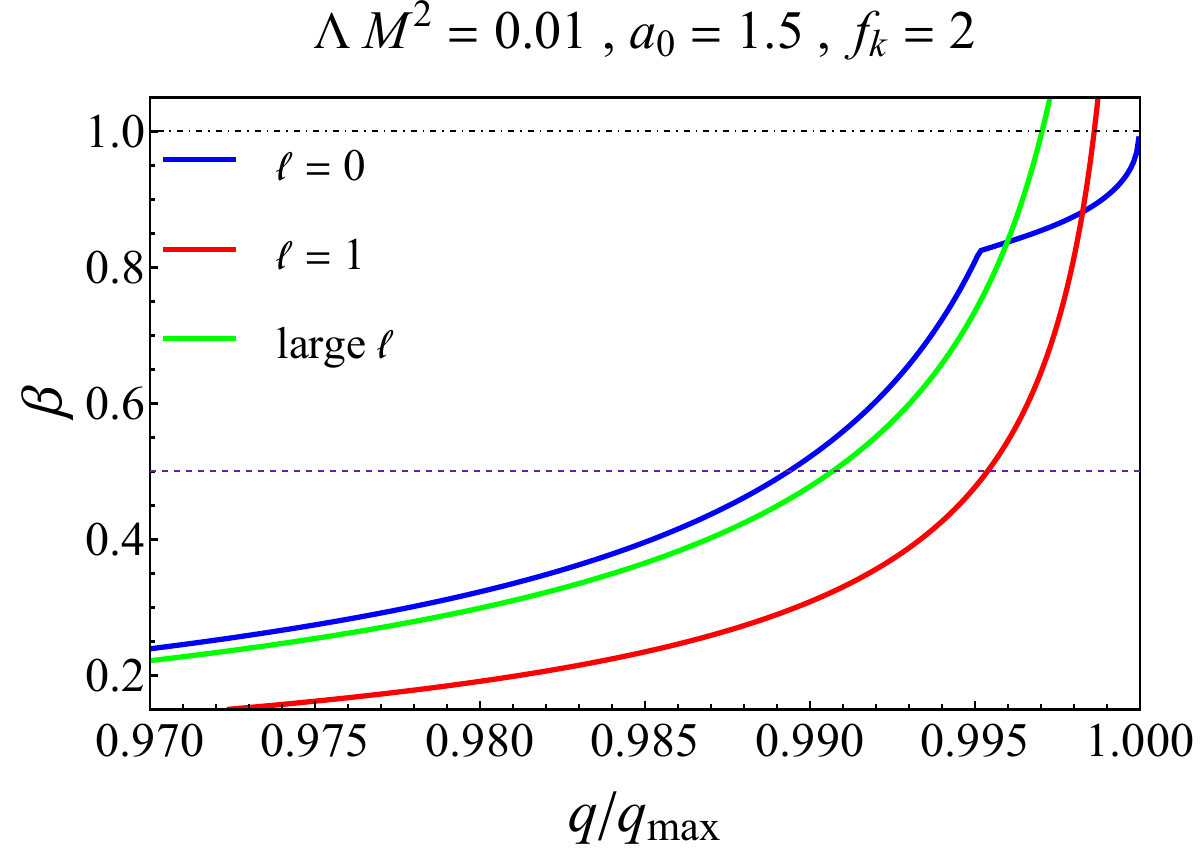}
   \end{minipage}
   }
    \subfigure{
   \begin{minipage}[h]{0.315\linewidth}
   \centering
   \includegraphics[width=\linewidth]{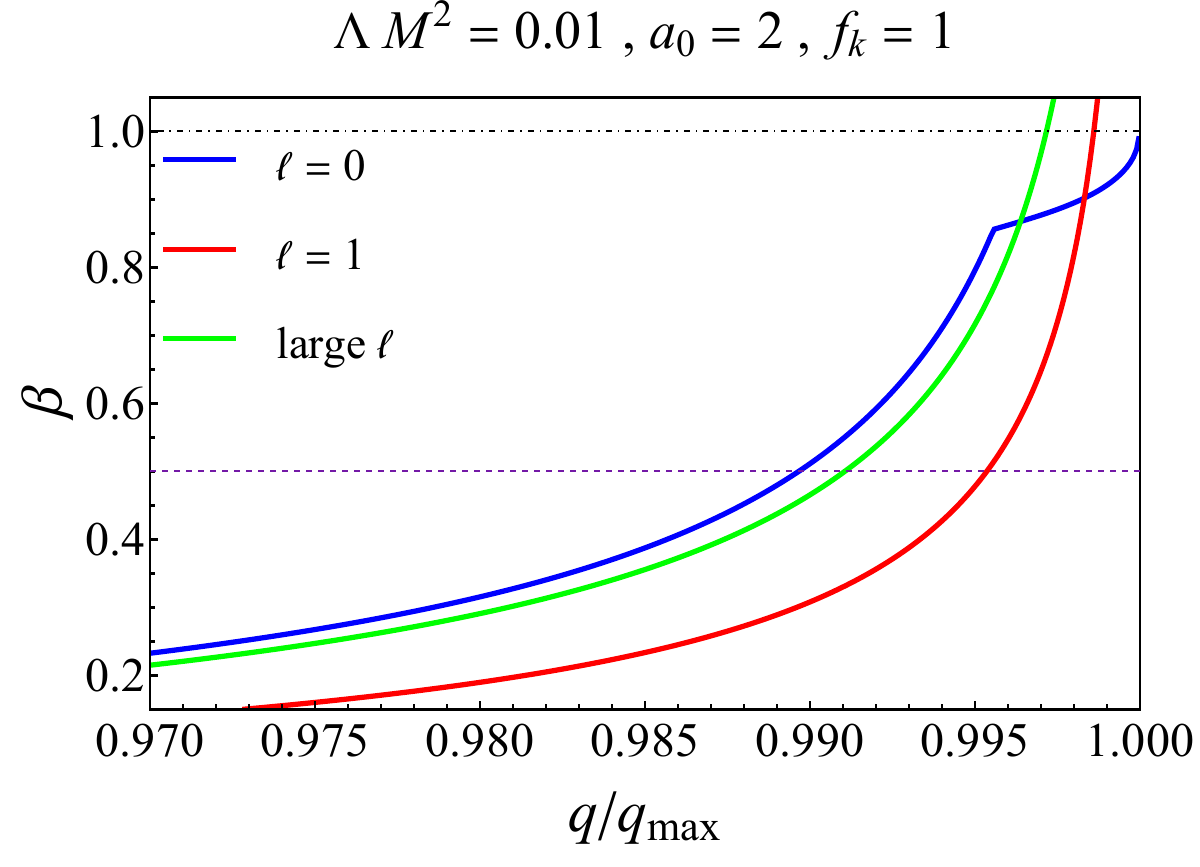}
   \end{minipage}
   }
    \subfigure{
   \begin{minipage}[h]{0.315\linewidth}
   \centering
   \includegraphics[width=\linewidth]{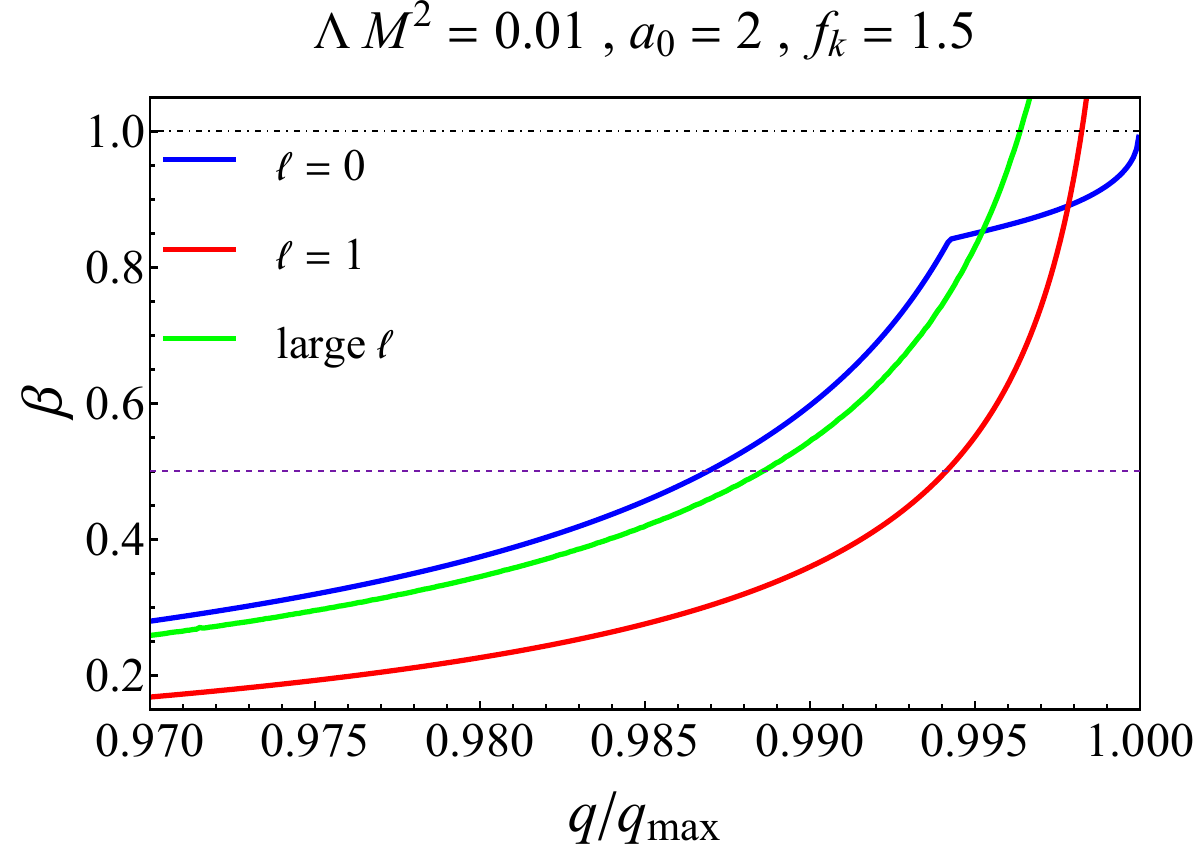}
   \end{minipage}
   }
   \subfigure{
   \begin{minipage}[h]{0.315\linewidth}
   \centering
   \includegraphics[width=\linewidth]{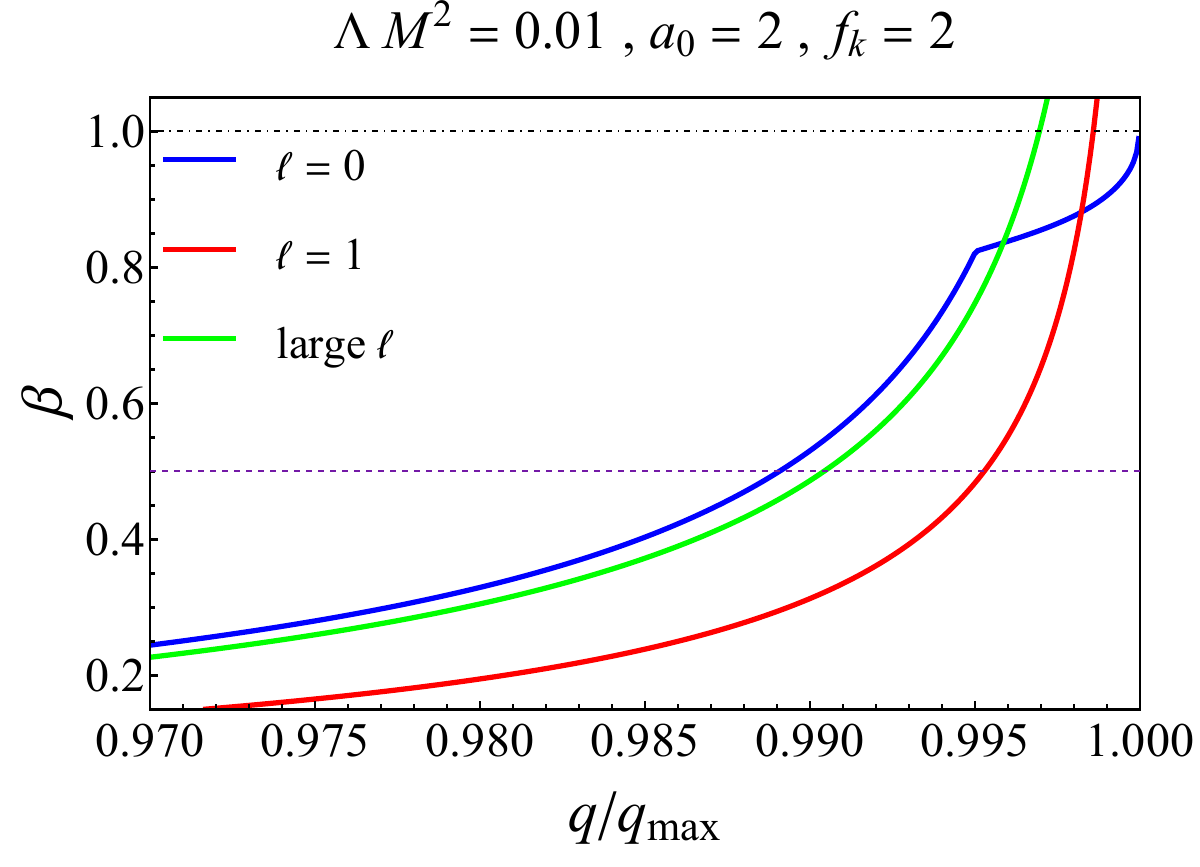}
   \end{minipage}
   }
   \caption{(Color online) Dominant modes $\beta$ varies with the charge ratio $q/q_{\rm max}$ for different parameters $a_0$ and $f_k$ when $\Lambda M^2=0.01$. The blue curve ($\ell=0$), red curve ($\ell=1$), green curve (large $\ell$) correspond to NE mode, dS mode, PS mode, respectively. The horizontal dashed line $\beta=1/2$ represents the threshold value for the SCC.}
    \label{fig:betaqqmaxlambda001}
\end{figure*}

\begin{figure*}[htbp]
   \centering
   \subfigure{
   \begin{minipage}[h]{0.315\linewidth}
   \centering
   \includegraphics[width=\linewidth]{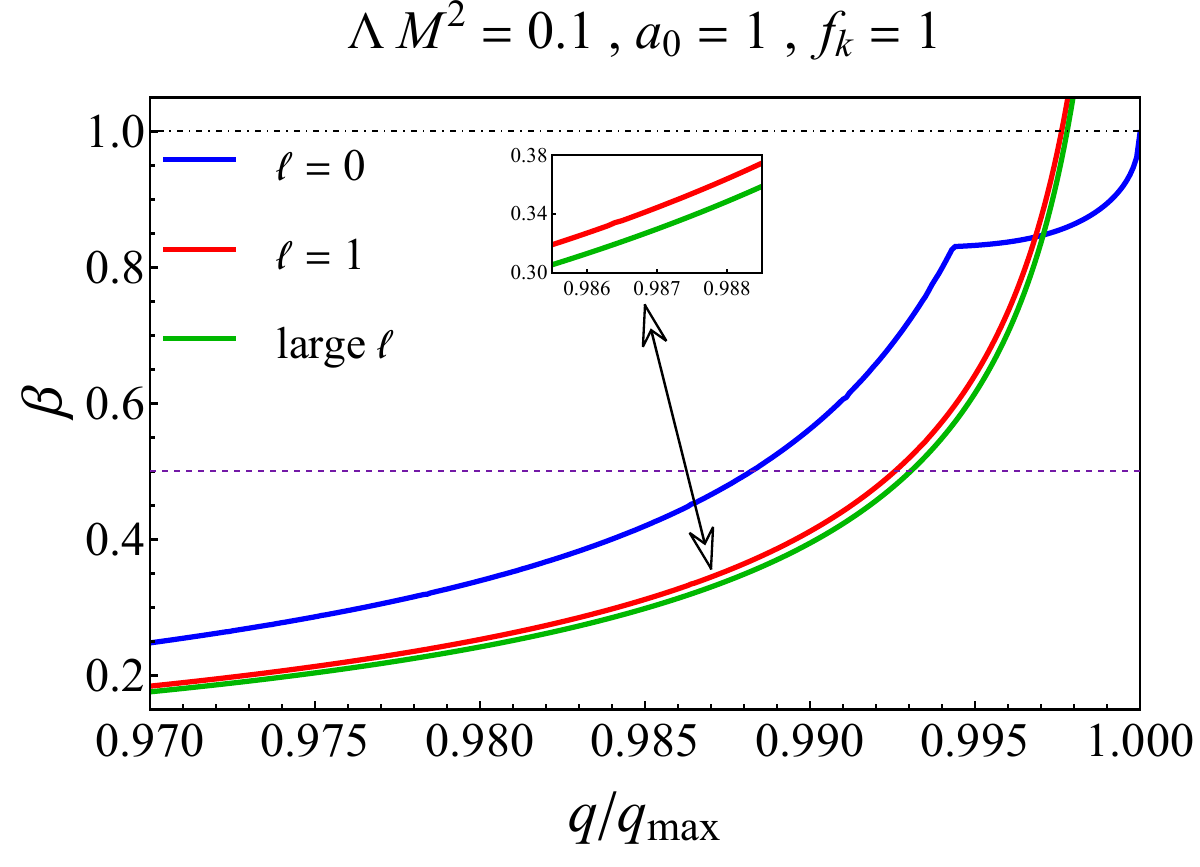}
   \end{minipage}
   }
    \subfigure{
   \begin{minipage}[h]{0.315\linewidth}
   \centering
   \includegraphics[width=\linewidth]{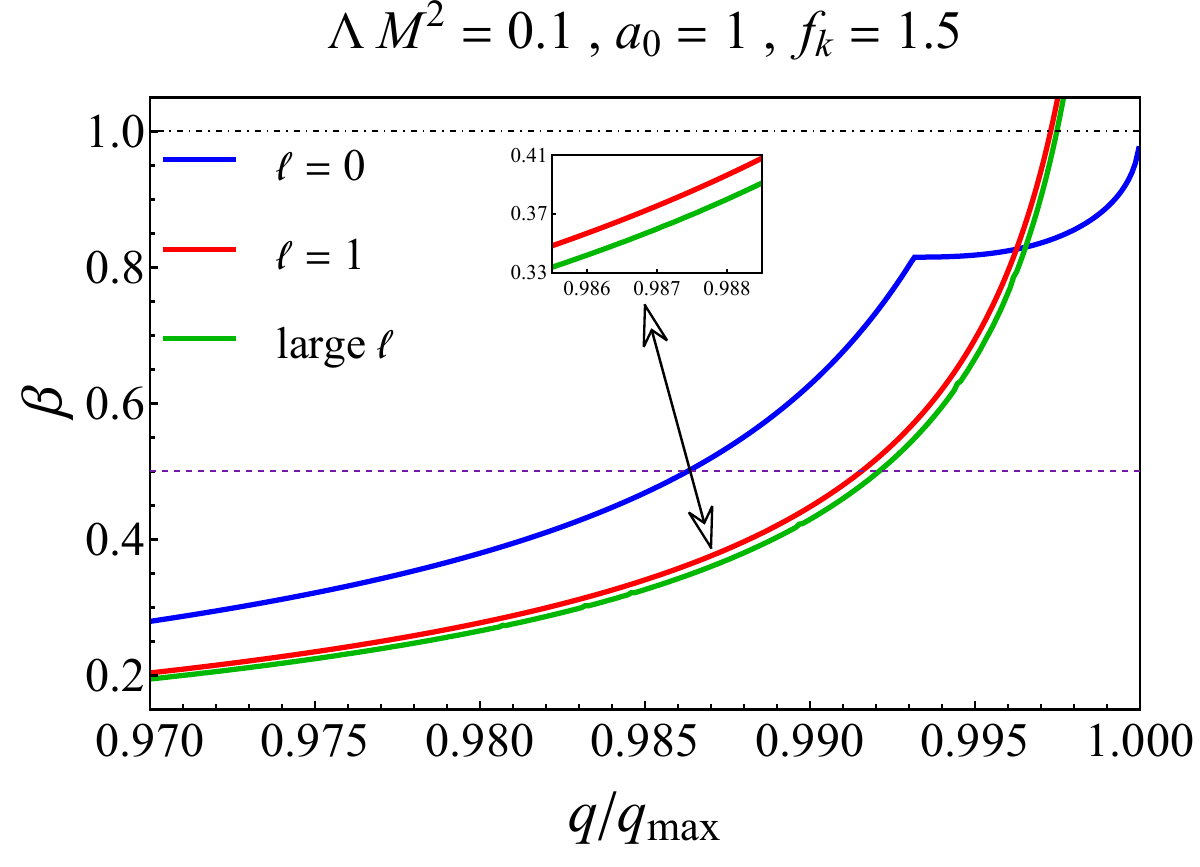}
   \end{minipage}
   }
   \subfigure{
   \begin{minipage}[h]{0.315\linewidth}
   \centering
   \includegraphics[width=\linewidth]{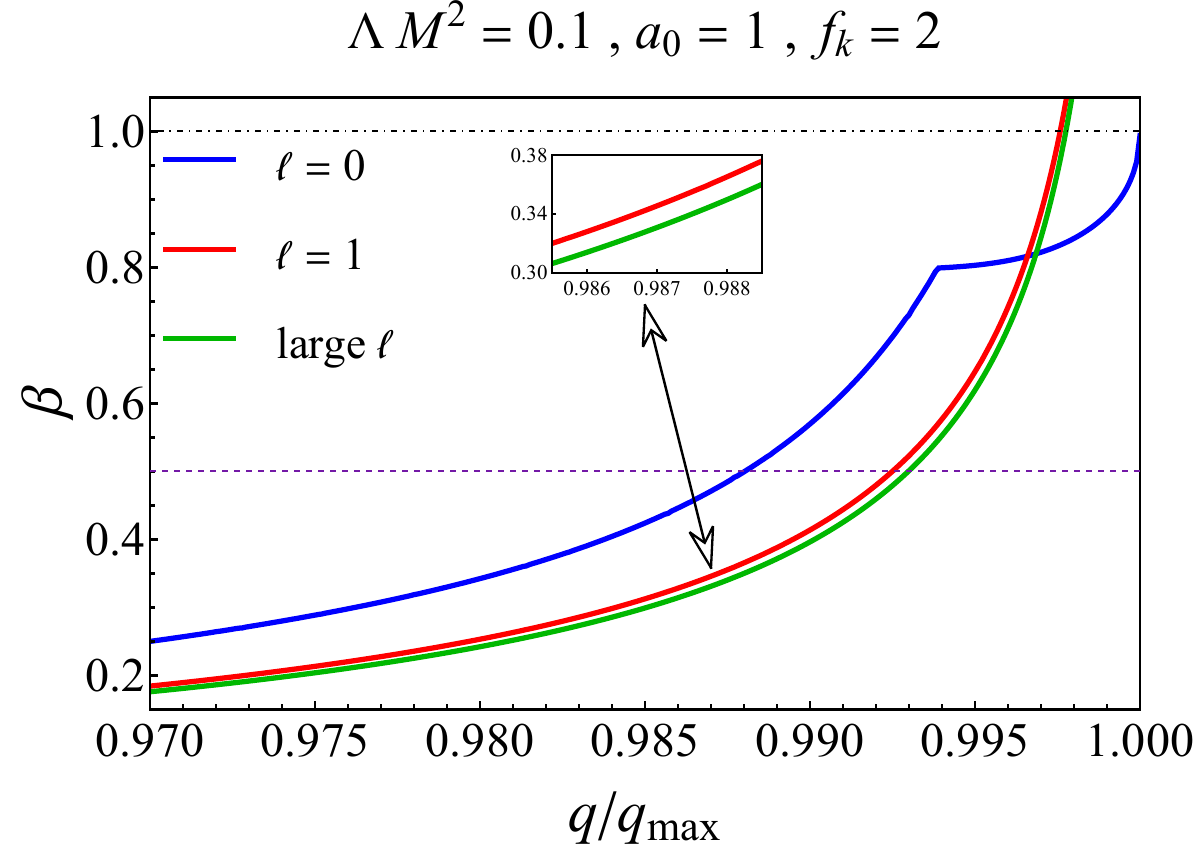}
   \end{minipage}
   }
    \subfigure{
   \begin{minipage}[h]{0.315\linewidth}
   \centering
   \includegraphics[width=\linewidth]{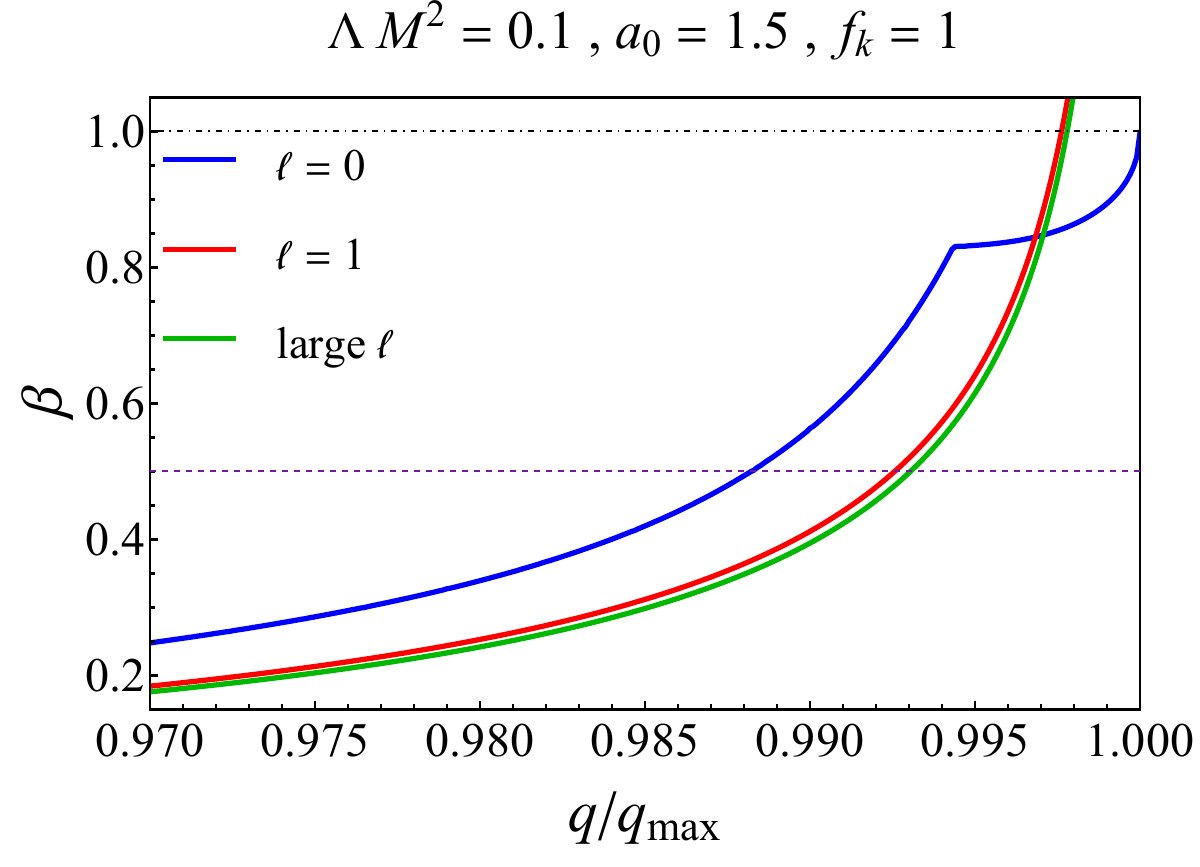}
   \end{minipage}
   }
    \subfigure{
   \begin{minipage}[h]{0.315\linewidth}
   \centering
   \includegraphics[width=\linewidth]{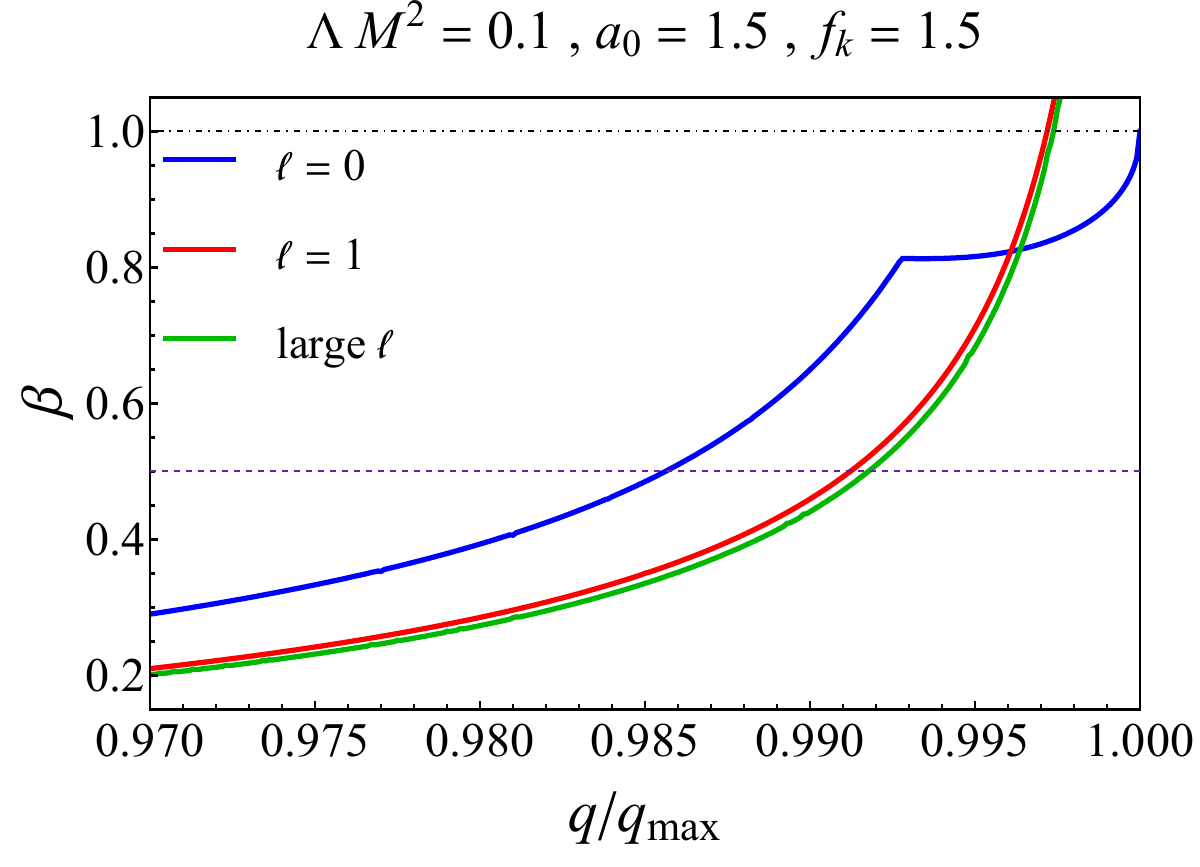}
   \end{minipage}
   }
   \subfigure{
   \begin{minipage}[h]{0.315\linewidth}
   \centering
   \includegraphics[width=\linewidth]{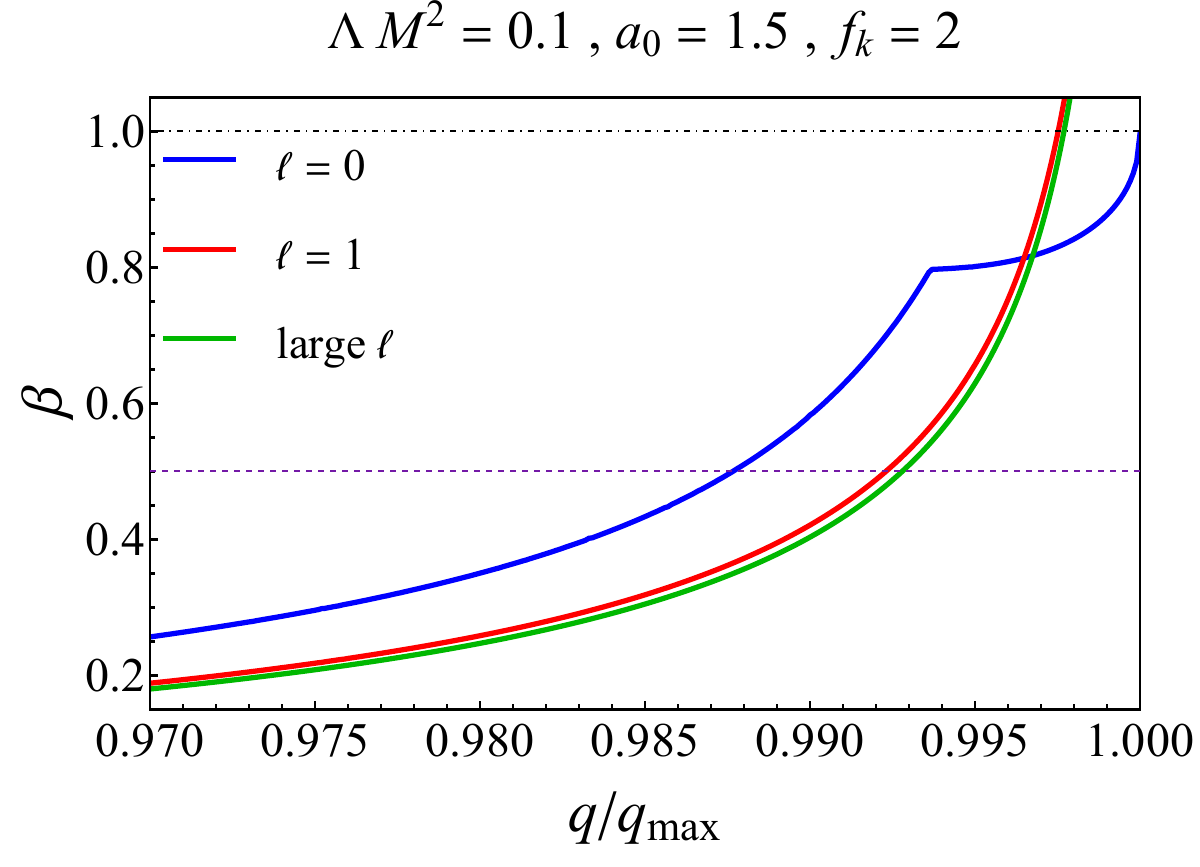}
   \end{minipage}
   }
    \subfigure{
   \begin{minipage}[h]{0.315\linewidth}
   \centering
   \includegraphics[width=\linewidth]{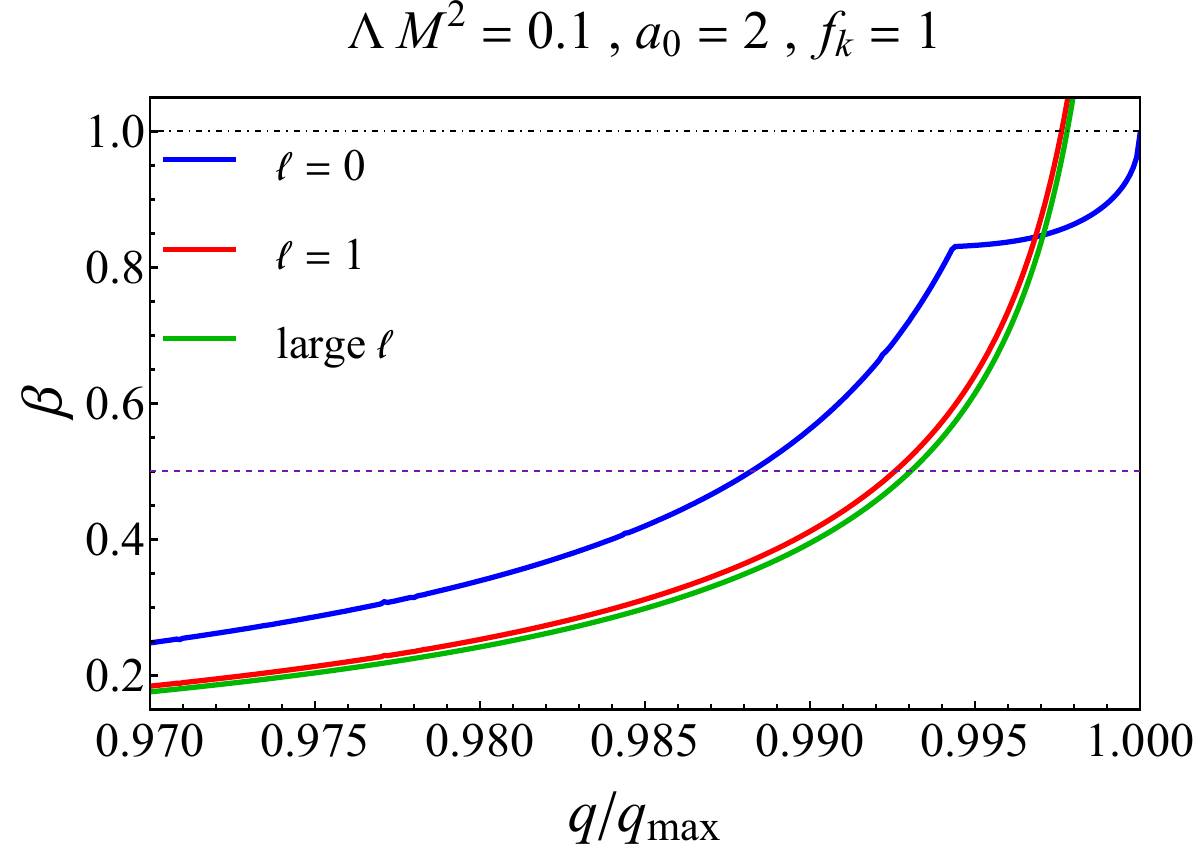}
   \end{minipage}
   }
    \subfigure{
   \begin{minipage}[h]{0.315\linewidth}
   \centering
   \includegraphics[width=\linewidth]{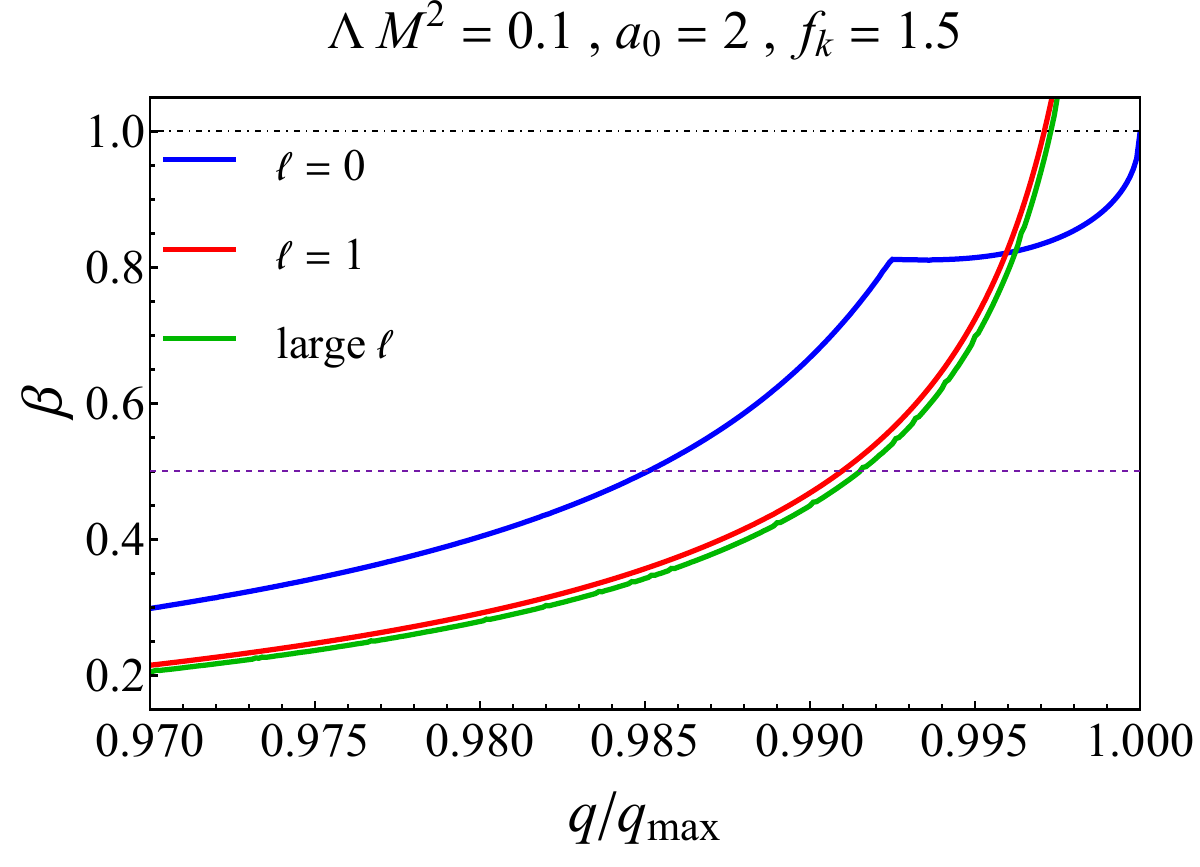}
   \end{minipage}
   }
   \subfigure{
   \begin{minipage}[h]{0.315\linewidth}
   \centering
   \includegraphics[width=\linewidth]{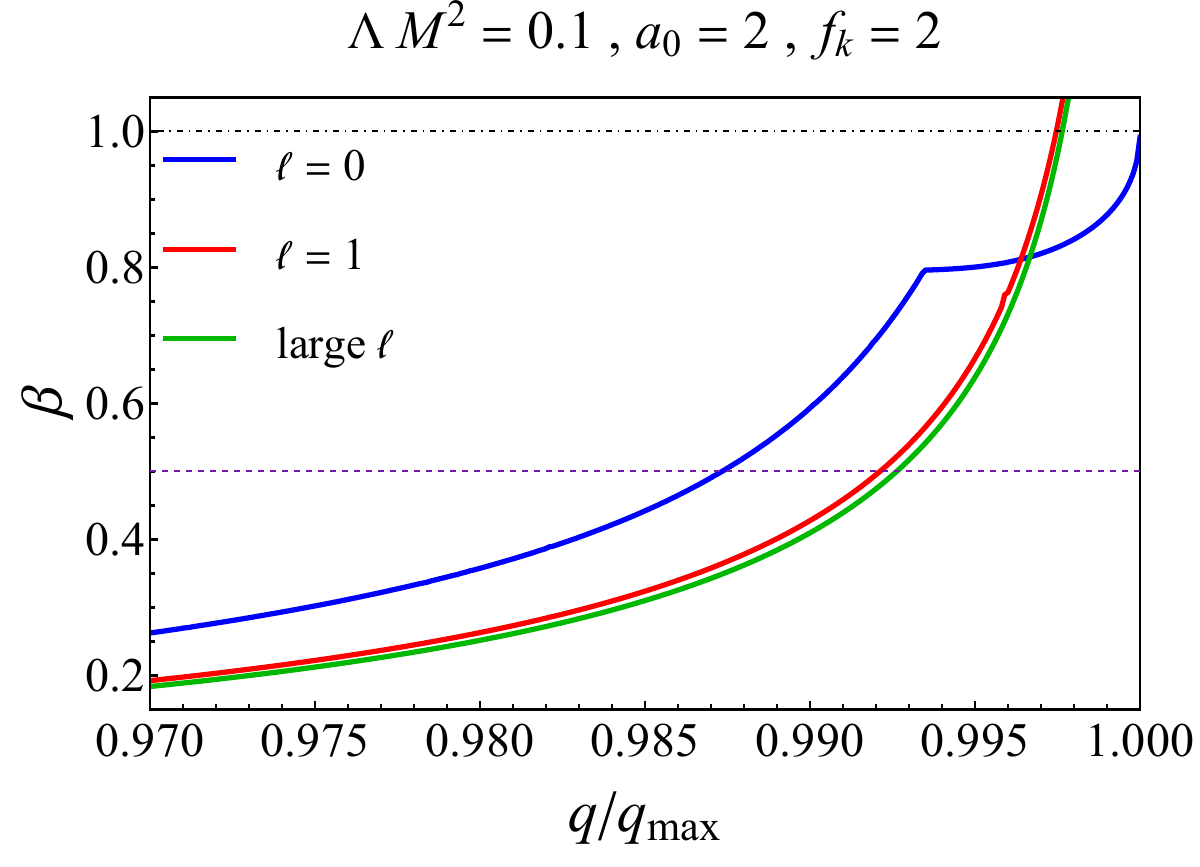}
   \end{minipage}
   }
   \caption{(Color online) Dominant modes $\beta$ varies with the charge ratio $q/q_{\rm max}$ for different parameters $a_0$ and $f_k$ when $\Lambda M^2=0.1$. The blue curve ($\ell=0$), red curve ($\ell=1$), green curve (large $\ell$) correspond to NE mode, dS mode, PS mode, respectively. The horizontal dashed line $\beta=1/2$ represents the threshold value for the SCC.}
    \label{fig:betaqqmaxlambda01}
\end{figure*}

To examine the fate of the SCC in the novel CKG theory, we further investigate the dominant modes. The evolution behaviors of $\beta$ in the parameter space can be witnessed in Fig.~\ref{fig:betaqqmaxlambda001} for cosmological constant $\Lambda M^2$$=$$0.01$ and Fig.~\ref{fig:betaqqmaxlambda01} for cosmological constant $\Lambda M^2$$=$$0.1$. The charge ratio $q/q_{\rm max}$$=$$1$ denotes the case of extremal black hole. We divide the QNMs into three families \cite{Cardoso:2017soq} for near-extremal black holes. The blue curve ($\ell$$=$$0$), red curve ($\ell$$=$$1$) and green curve (large $\ell$) correspond to near extremal (NE) mode, de Sitter (dS) mode and photon sphere (PS) mode, respectively. The dominant modes are defined by lowest-lying branches. Seen from Fig.~\ref{fig:betaqqmaxlambda001} for a small cosmological constant $\Lambda M^2$$=$$0.01$, the lowest-lying modes consist of two different branches. The lowest-lying dS mode ($\ell$$=$$1$) for the far-from-extremal black hole and the NE mode ($\ell$$=$$0$) for the black hole approaching extremality together constitute the dominant modes. As the cosmological constant increases to $\Lambda M^2$$=$$0.1$ showing in Fig.~\ref{fig:betaqqmaxlambda01}, it is quite interesting that the PS mode (large $\ell$) replaces the dS mode as the lowest-lying branch for the far-from-extremal black hole. We can see the closeup in the top three subfigures of Fig.~\ref{fig:betaqqmaxlambda01} for a vivid illustration. Ultimately, the lowest-lying PS mode and the NE mode together serve as the dominant modes for relatively large $\Lambda M^2$. As a result, we find that the dominant modes 
exhibit a strong dependence on the cosmological constant $\Lambda M^2$, especially for the branch of the far-from-extremal black hole.

Observing these two figures closely, it is notable that the dominant mode branches rise monotonically with the charge ratio $q/q_{\rm max}$ increasing, and eventually cross over the threshold line $\beta$$=$$1/2$ for near-extremal black holes. It indicates that the SCC is violated. Consequently, under the neutral massless scalar field perturbation, the SCC cannot yet gets rid off the fate of being violated in the framework of CKG theory although the NLED effect is also taken into account.

\begin{figure*}[htbp]
   \centering
   \subfigure{
   \begin{minipage}[h]{0.315\linewidth}
   \centering
   \includegraphics[width=\linewidth]{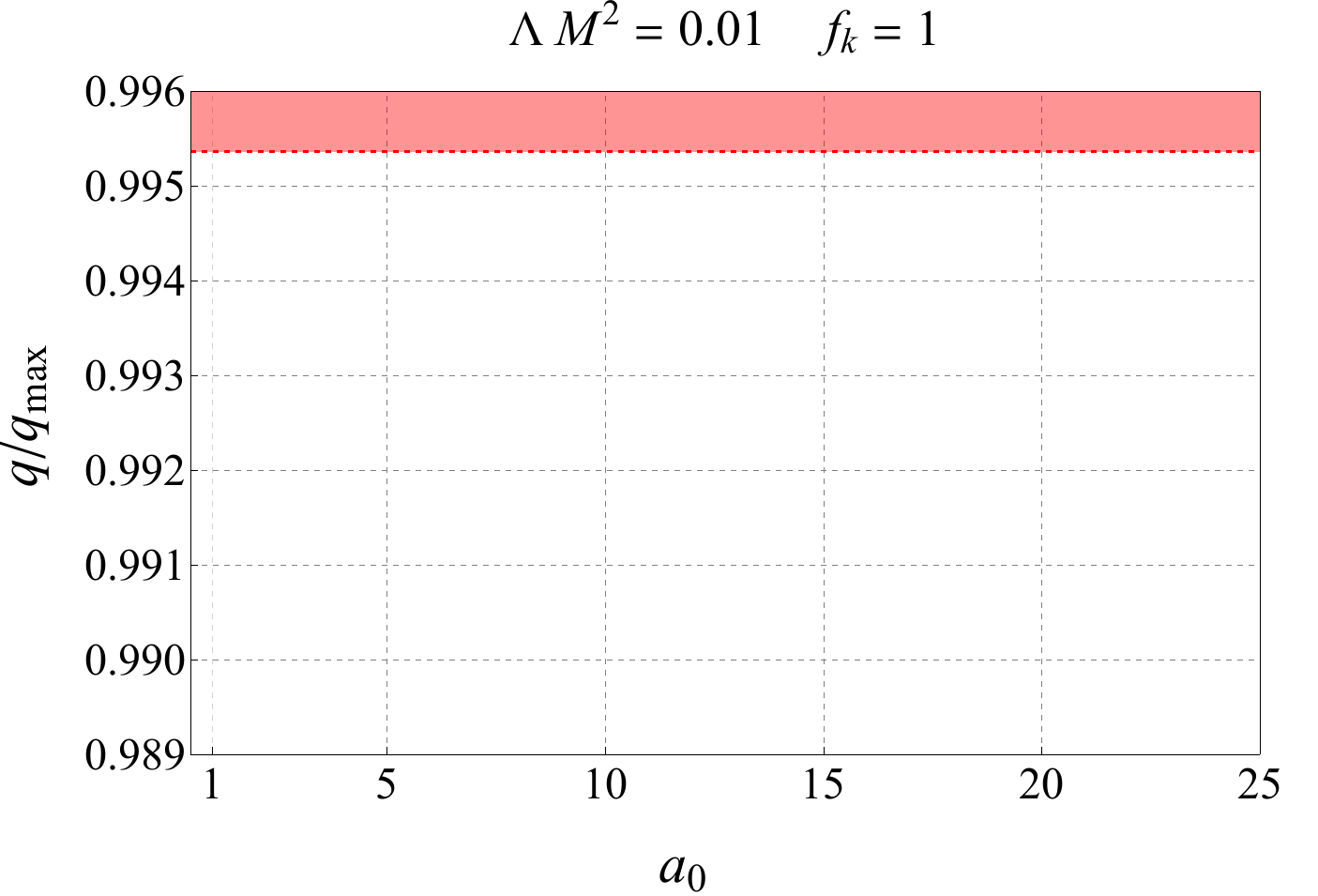}
   \end{minipage}
   }
    \subfigure{
   \begin{minipage}[h]{0.315\linewidth}
   \centering
   \includegraphics[width=\linewidth]{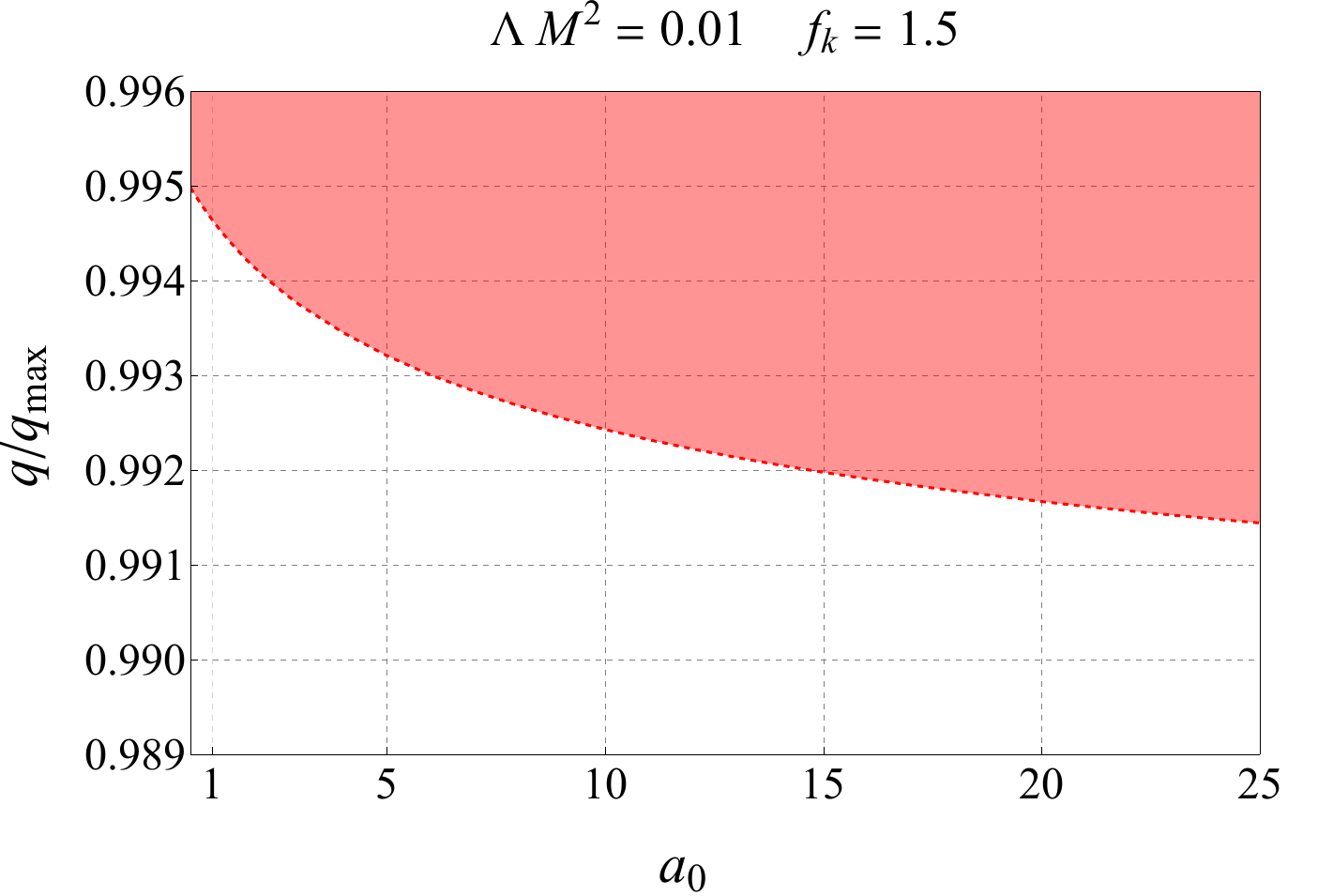}
   \end{minipage}
   }
   \subfigure{
   \begin{minipage}[h]{0.315\linewidth}
   \centering
   \includegraphics[width=\linewidth]{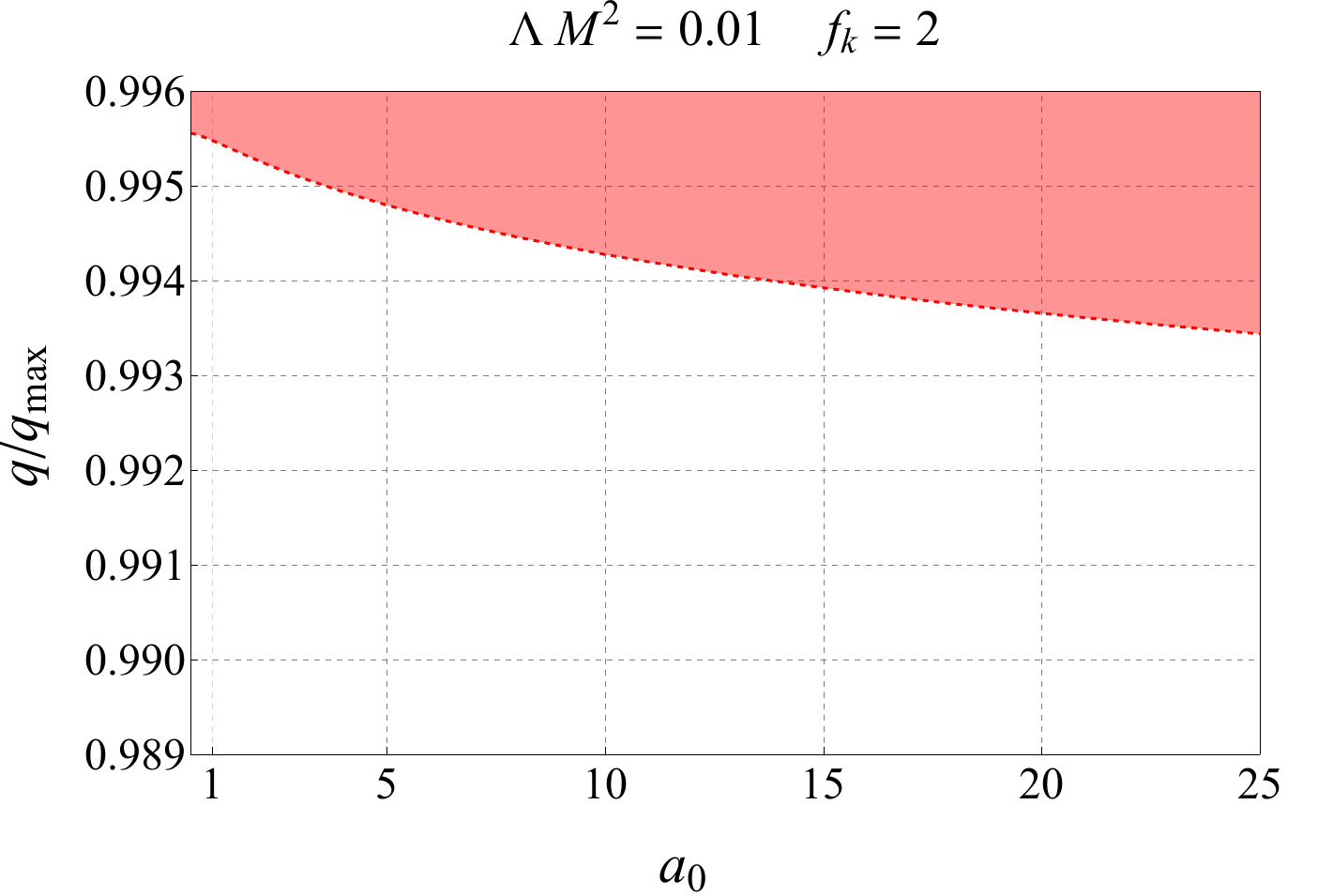}
   \end{minipage}
   }
    \subfigure{
   \begin{minipage}[h]{0.315\linewidth}
   \centering
   \includegraphics[width=\linewidth]{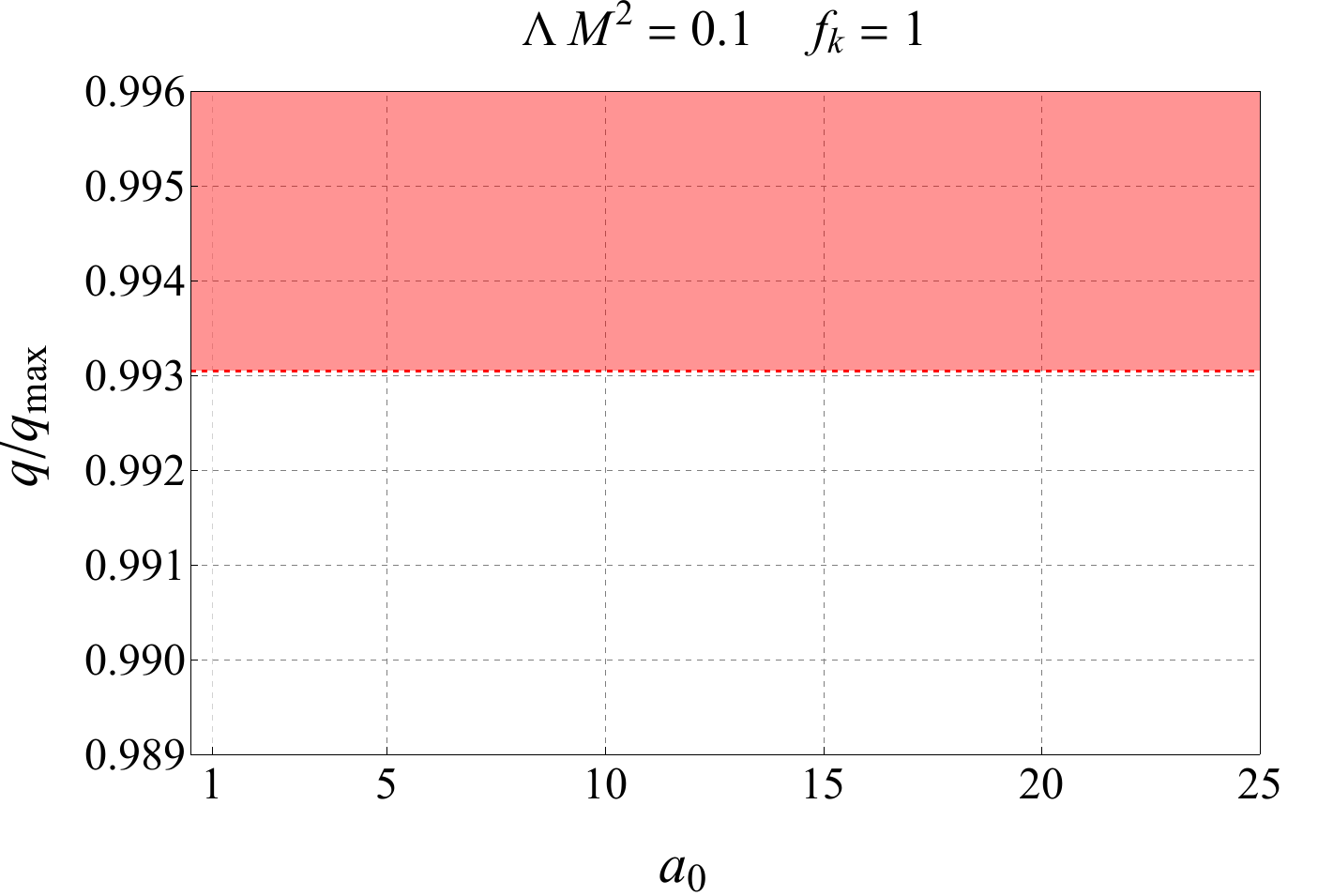}
   \end{minipage}
   }
    \subfigure{
   \begin{minipage}[h]{0.315\linewidth}
   \centering
   \includegraphics[width=\linewidth]{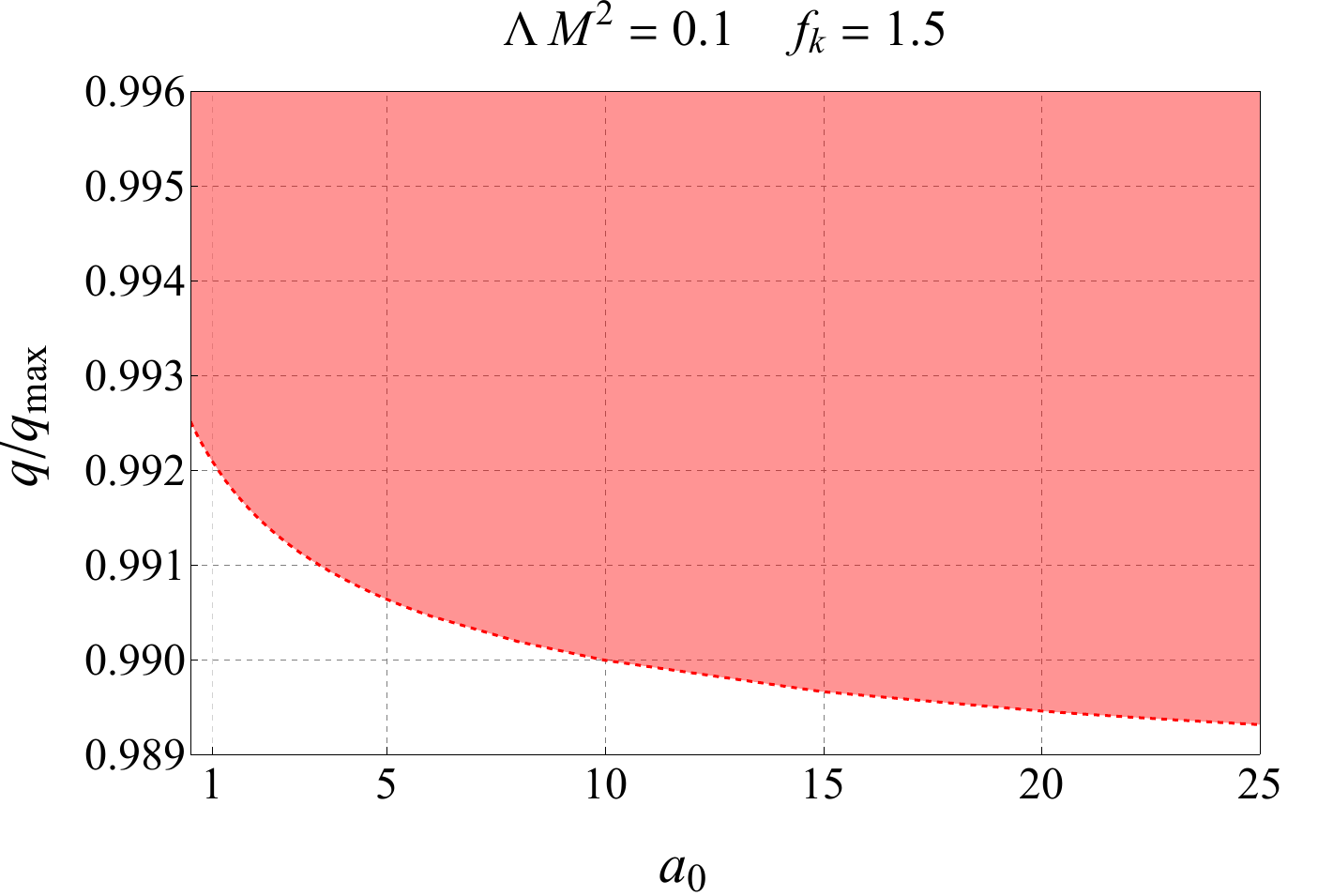}
   \end{minipage}
   }
   \subfigure{
   \begin{minipage}[h]{0.315\linewidth}
   \centering
   \includegraphics[width=\linewidth]{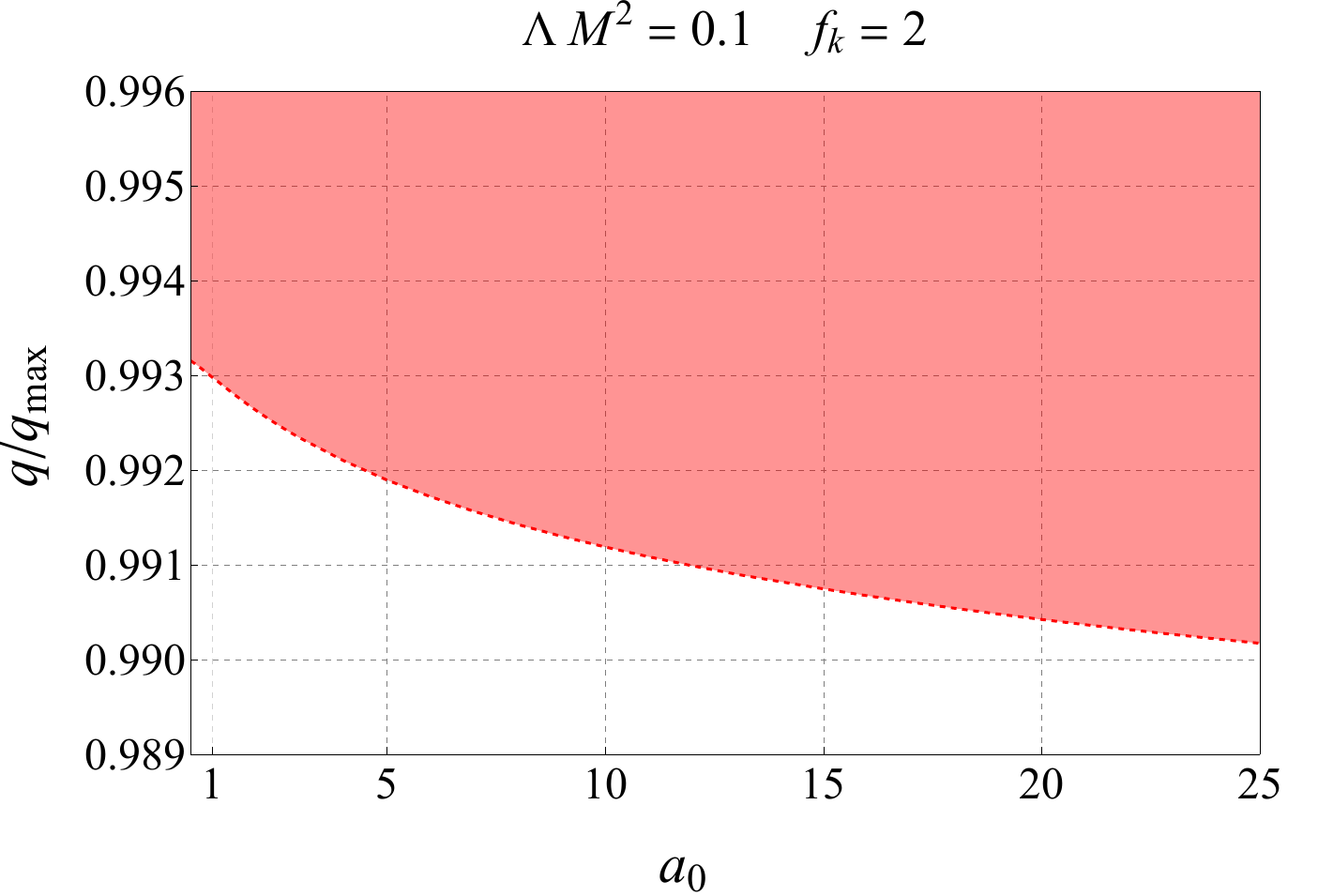}
   \end{minipage}
   }
   \caption{(Color online) The charge ratio $q/q_{\rm max}$ varies with $a_0$ in parameter space. The red shaded region signals the SCC violation.}
   \label{fig:qqmaxa0}
\end{figure*}

\begin{figure*}[htbp]
   \centering
   \subfigure{
   \begin{minipage}[h]{0.315\linewidth}
   \centering
   \includegraphics[width=\linewidth]{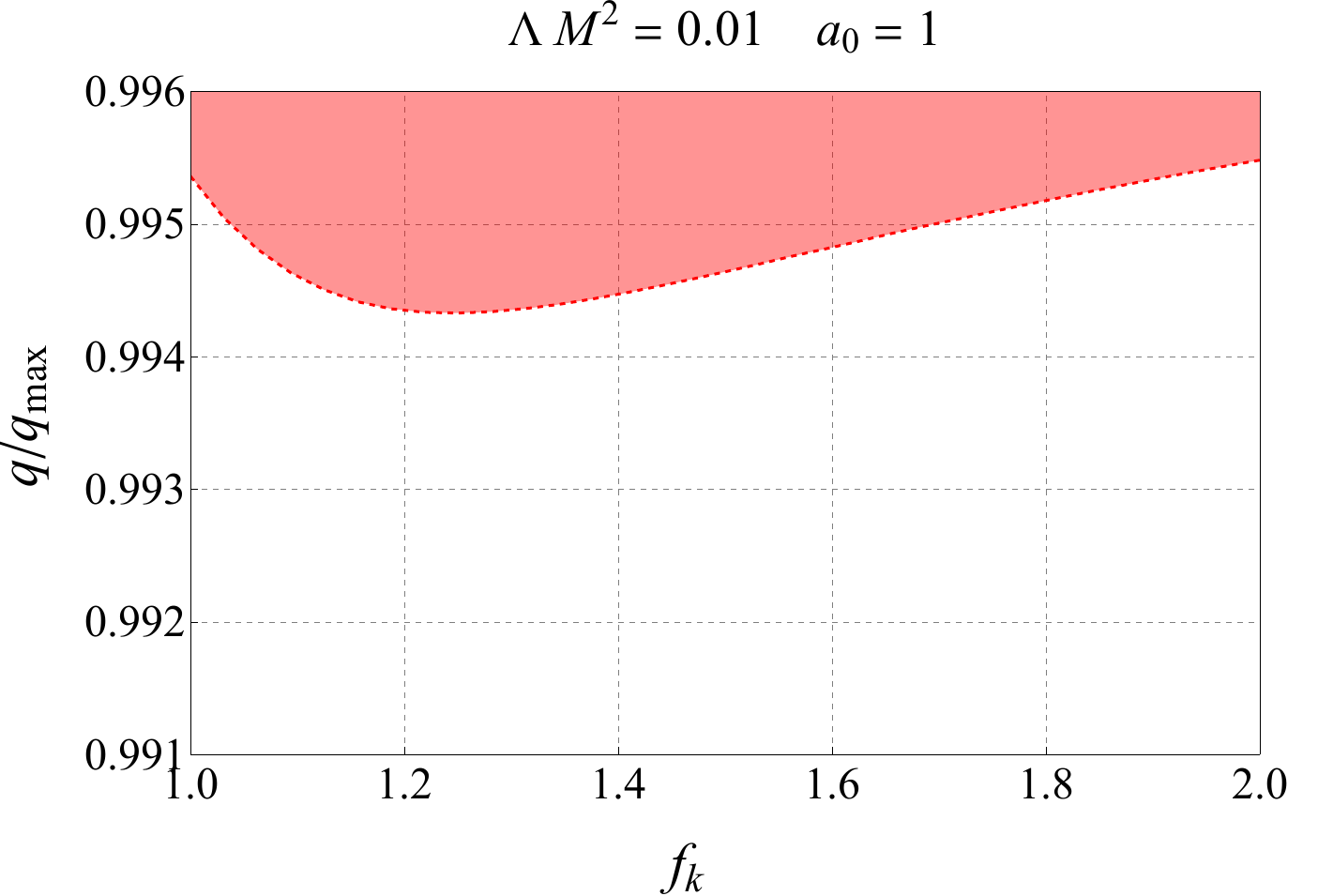}
   \end{minipage}
   }
    \subfigure{
   \begin{minipage}[h]{0.315\linewidth}
   \centering
   \includegraphics[width=\linewidth]{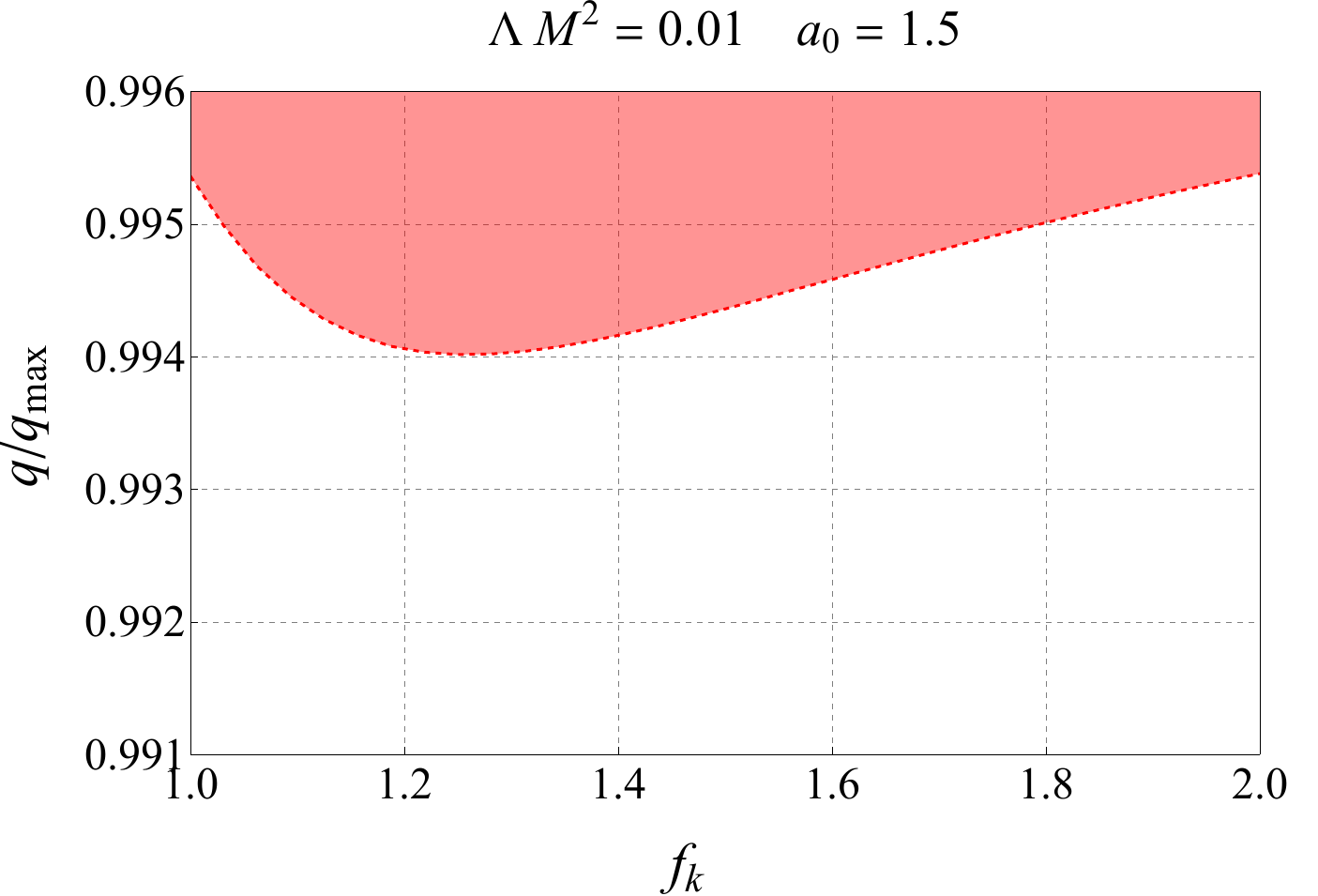}
   \end{minipage}
   }
   \subfigure{
   \begin{minipage}[h]{0.315\linewidth}
   \centering
   \includegraphics[width=\linewidth]{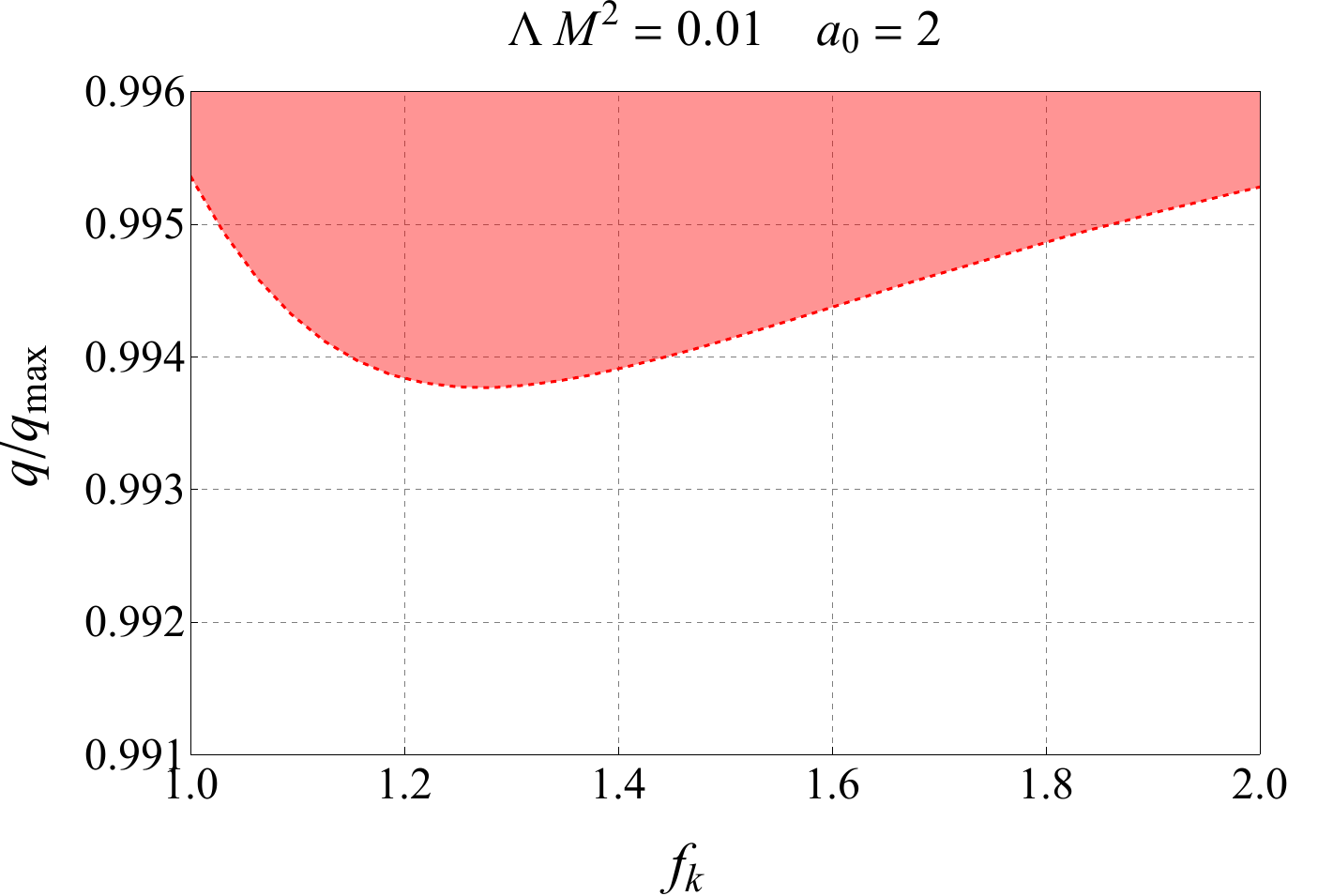}
   \end{minipage}
   }
    \subfigure{
   \begin{minipage}[h]{0.315\linewidth}
   \centering
   \includegraphics[width=\linewidth]{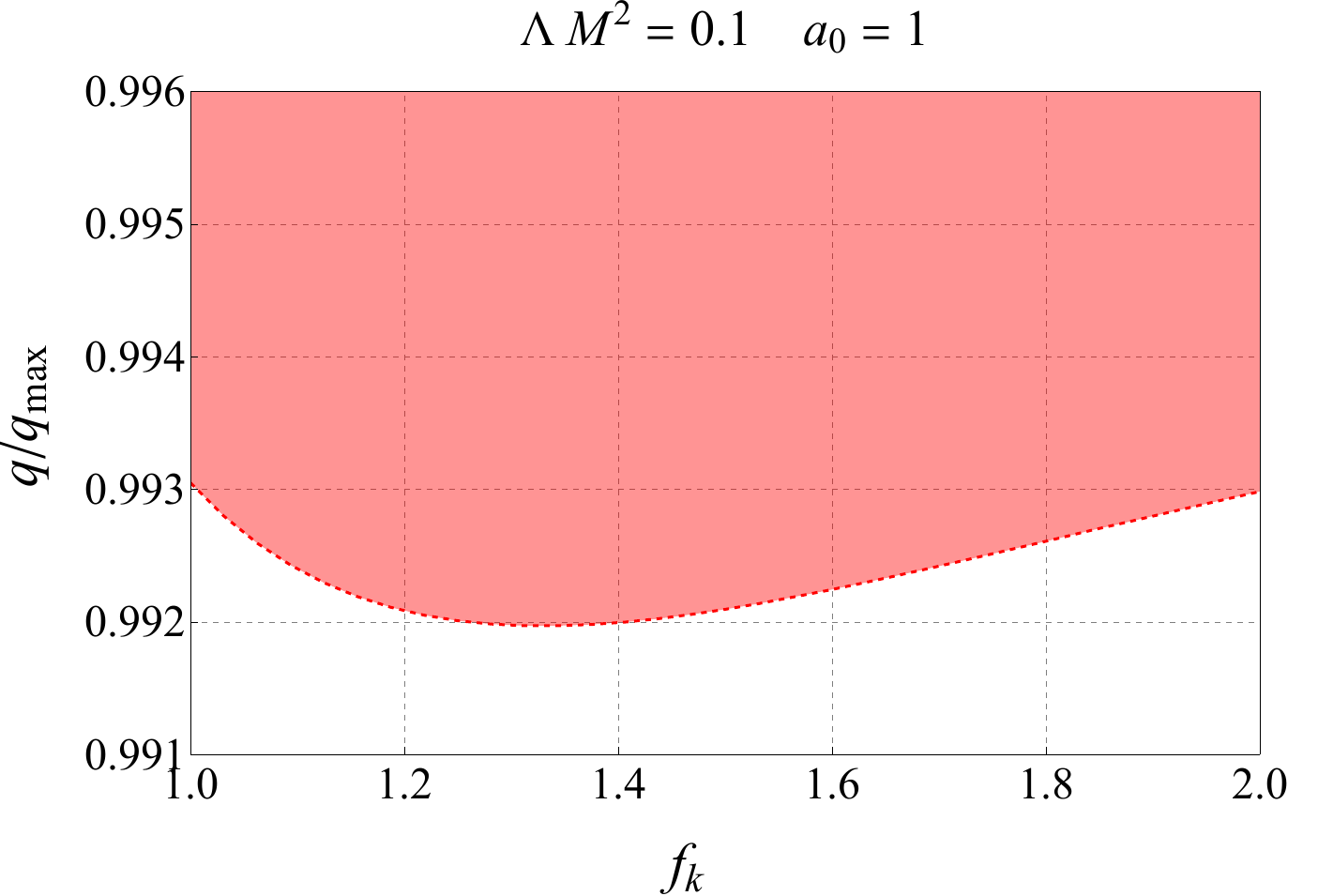}
   \end{minipage}
   }
    \subfigure{
   \begin{minipage}[h]{0.315\linewidth}
   \centering
   \includegraphics[width=\linewidth]{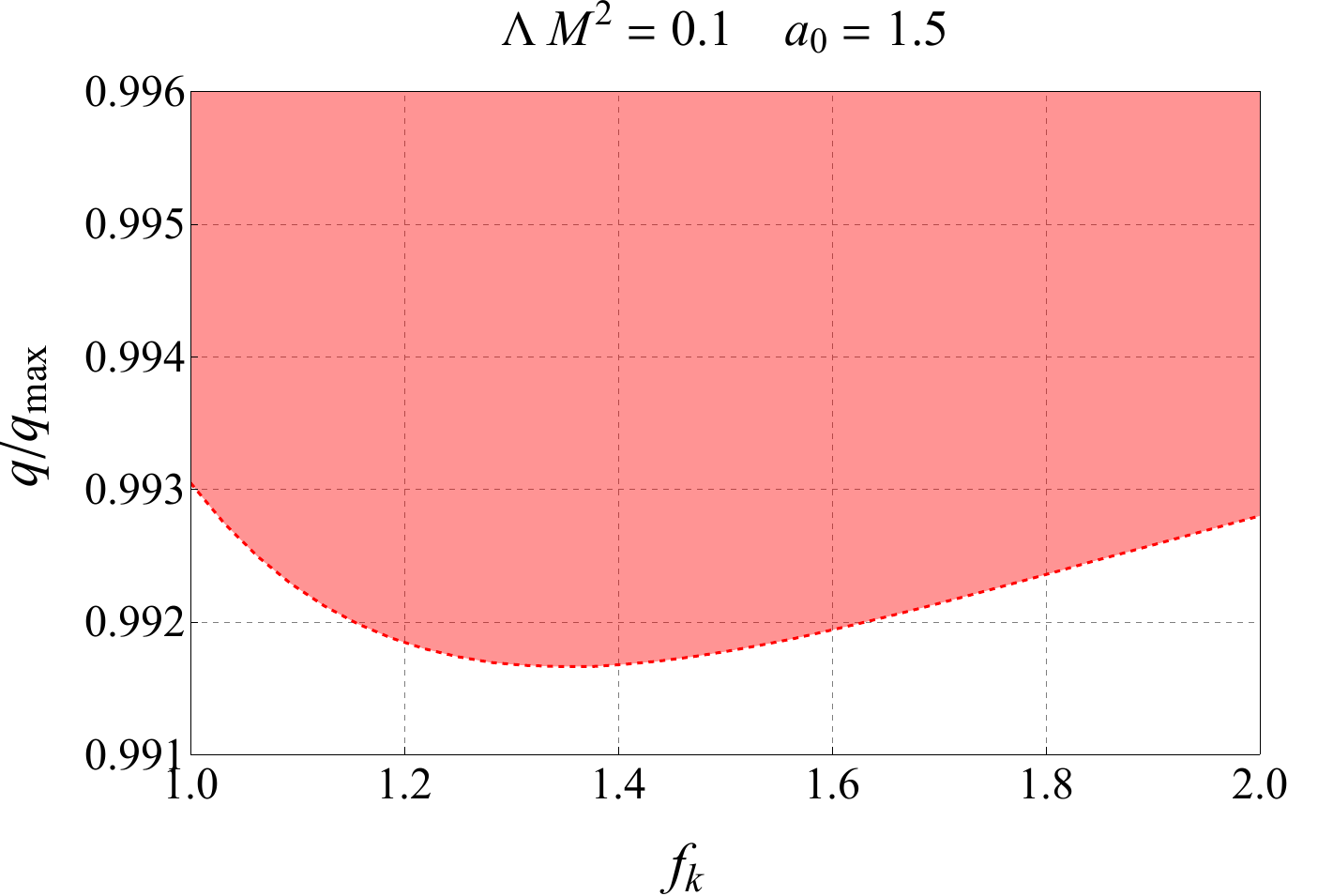}
   \end{minipage}
   }
   \subfigure{
   \begin{minipage}[h]{0.315\linewidth}
   \centering
   \includegraphics[width=\linewidth]{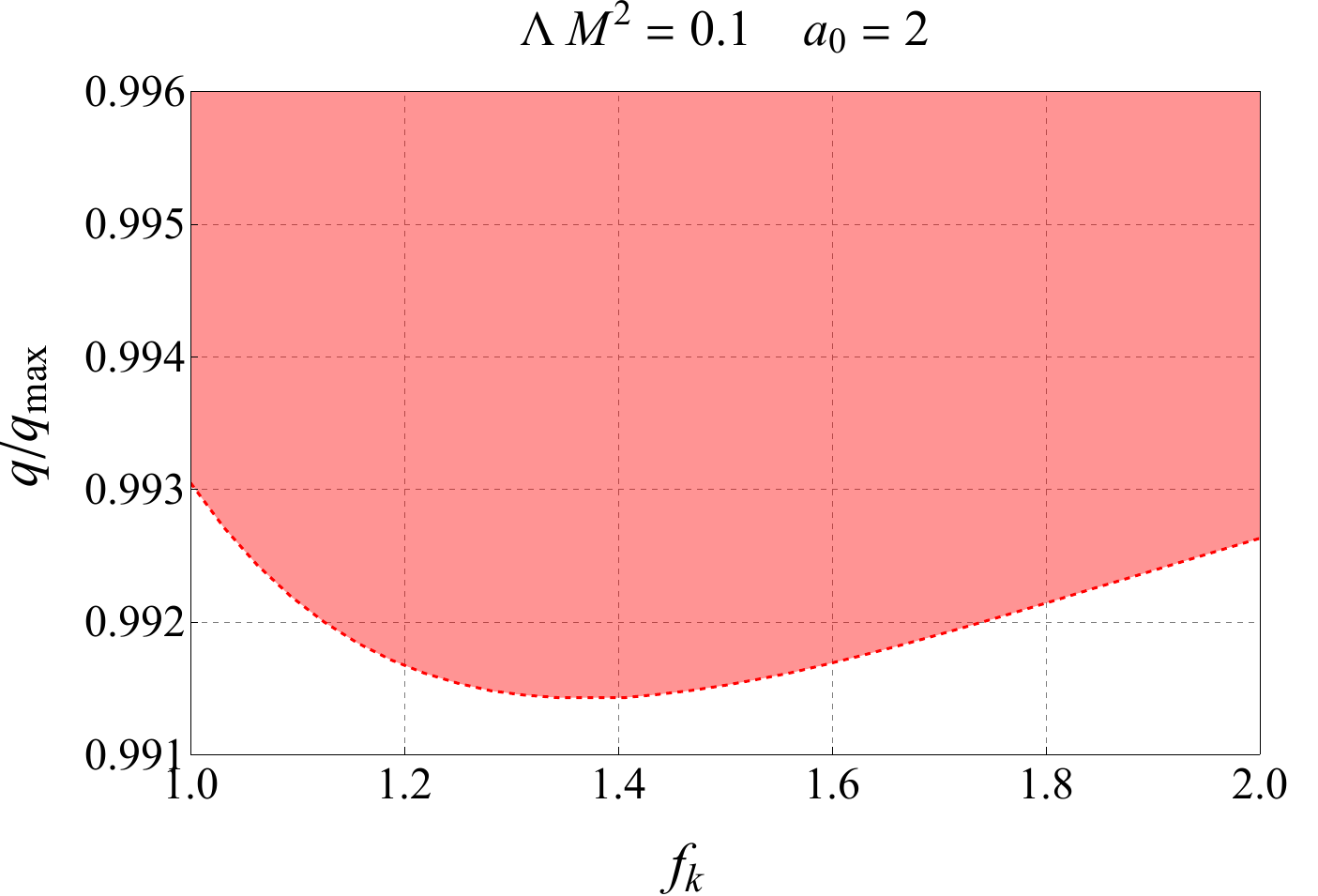}
   \end{minipage}
   }
   \caption{(Color online) The charge ratio $q/q_{\rm max}$ varies with $f_k$ in parameter space. The red shaded region signals the SCC violation.}
   \label{fig:qqmaxfk}
\end{figure*}

Next, we systematically analyze how the key characteristic parameters, including the CKG and NLED modifications, as well as the cosmological constant, affect the validity of the SCC. As illustrated in Fig.~\ref{fig:qqmaxa0} and Fig.~\ref{fig:qqmaxfk}, the charge ratio $q/q_{\rm max}$ varies with parameters $a_0$ and $f_k$ for different cosmological constant $\Lambda M^2$. The subfigures in each column correspond to the same $f_k$ with $\Lambda M^2$ increasing from the top to the bottom. The red shaded region signals the SCC violation. Seen from the Fig.~\ref{fig:qqmaxa0}, the violation region expands initially and then decreases with the parameter $f_k$ increasing from the left to the right. The non-monotonic effects caused by the parameter $f_k$ can also be captured in Fig.~\ref{fig:qqmaxfk}. What's more, as the parameter $a_0$ increases, Fig.~\ref{fig:qqmaxfk} shows a steady expansion of the violation region, implying that the parameter $a_0$ may promote the de Sitter black hole to violate the SCC. A careful comparison between the top three subfigures and the bottom ones in Fig.~\ref{fig:qqmaxa0} or Fig.~\ref{fig:qqmaxfk} reveals that the cosmological constant $\Lambda M^2$ plays a role analogous to the parameter $a_0$ with respect to the validity of the SCC. Therefore, our analyses demonstrate that both the parameter $a_0$ and the cosmological constant $\Lambda M^2$ facilitate the violation of the SCC, whereas the parameter $f_k$ causes the violation region to expand initially and then decrease.

\section{Conclusions} \label{SecCon}

The confidence in SCC has been challenged since the discovery of some counterexamples in Einstein's theory, particularly for the de Sitter spacetime. In this paper, we extended to investigate the QNMs and SCC of the novel charged de Sitter black hole in the framework of CKG theory, thereby filling the gap in SCC research within the CKG framework.

Considering the neutral massless scalar perturbations, we first derived the criterion to measure the validity of the SCC. Of special interest, the threshold for the criterion in the CKG theory takes the same value as that for Reissner-Nordstr{\"o}m de Sitter black holes in Einstein's gravity. Numerical analyses mainly via the pseudospectral method reveal that the dominant QNMs always consist of two different branches. The branch for the far-from-extremal black hole exhibits a strong dependence on the cosmological constant $\Lambda M^2$. Unexpectedly, the fact that the dominant mode branches rise monotonically and eventually cross over the threshold line $\beta$$=$$1/2$ when the black hole approaches extremality manifests that the SCC remains violated. Consequently, the CKG and NLED modifications do not suffice to rescue the SCC. What's more, our further study demonstrates the pivotal influence of the characteristic parameters on the validity of the SCC. In detail, both the parameter $a_0$ and the cosmological constant $\Lambda M^2$ promote the SCC to be violated, while the parameter $f_k$ causes the violation region to expand initially and then decrease.

As the first instance for investigating SCC within the CKG theory, we have demonstrated that the SCC still suffers from violation for the de Sitter spacetime although the NLED effect was also considered. This paper not only provides new insights into SCC beyond Einstein's general relativity, but also contributes to deeper understanding this fascinating gravity theory.

\section*{Declaration of Competing Interest}
The authors declare that they have no known competing financial interests or personal relationships that could have appeared to influence the work reported in this paper.

\section*{Data Availability}
No data was used for the research described in the article.

\begin{acknowledgments}
The authors express sincere gratitude to Prof. Yi Ling for helpful discussions. This work is supported by the National Natural Science Foundation of China under grant No.11947086, the Sichuan Science and Technology Program under grant No.2022NSFSC1806, the Fundamental Scientific Research Foundation of China West Normal University under grant No.22kA005.
\end{acknowledgments}

\bibliographystyle{apsrev4-1}
\bibliography{QNMsSCCCKGRefs}

\end{document}